\documentclass[12pt]{article}

\usepackage[utf8]{inputenc}
\usepackage[english]{babel}
\usepackage{amsmath}
\usepackage{amssymb}
\usepackage{bbm}
\usepackage{hyperref}
\usepackage{feynmp-auto}
\usepackage{physics}
\usepackage{graphicx}

\usepackage{cite}
\usepackage{dcolumn}
\newcolumntype{+}{D{+}{\,\pm\,}{-3,3}}

\newcommand{\be}{\begin{equation}}
\newcommand{\ee}{\end{equation}}
\newcommand{\ba}{\begin{eqnarray}}
\newcommand{\ea}{\end{eqnarray}}
\newcommand{\bs}{\begin{subequations}}
\newcommand{\es}{\end{subequations}}
\newcommand{\no}{\nonumber\\}

\graphicspath{ {figures/} }

\begin{document}

\title{\LARGE On the addition of a large scalar multiplet
to the Standard Model}

\author{
  Darius~Jur\v{c}iukonis$^{(1)}$\thanks{E-mail:
    \tt darius.jurciukonis@tfai.vu.lt}
  \
  and Lu\'\i s~Lavoura$^{(2)}$\thanks{E-mail:
    \tt balio@cftp.tecnico.ulisboa.pt}
  \\*[3mm]
  $^{(1)}\!$
  \small Vilnius University, Institute of Theoretical Physics and Astronomy, \\
  \small Saul\.etekio~av.~3, Vilnius 10257, Lithuania
  \\*[2mm]
  $^{(2)}\!$
  \small Universidade de Lisboa, Instituto Superior T\'ecnico, CFTP, \\
  \small Av.~Rovisco~Pais~1, 1049-001~Lisboa, Portugal
}

\maketitle

\begin{abstract}
  We consider the addition of a single $SU(2)$ multiplet
  of complex scalar fields to the Standard Model (SM).
  We explicitly consider the various possible values of the weak isospin $J$
  of that multiplet,
  up to and including $J = 7/2$.
  We allow the multiplet to have arbitrary weak hypercharge.
  The scalar fields of the multiplet
  are assumed to have no vacuum expectation value;
  the mass differences among the components of the multiplet
  originate in its coupling,
  present in the scalar potential (SP),
  to the Higgs doublet of the SM.
  We derive exact bounded-from-below and unitarity conditions on the SP,
  thereby constraining those mass differences.
  We compare those constraints to the ones
  that may be derived from the oblique parameters.
\end{abstract}

\section{Introduction}

In this paper,
we study the model of New Physics (NP),
\textit{i.e.}~of physics beyond the Standard Model (SM),
wherein one adds to the SM one gauge-$SU(2)$ multiplet $\chi$
with weak isospin $J$
and consisting of $n = 2 J + 1$ complex scalar fields.
The multiplet has un-specified weak hypercharge $Y$;
therefore,
the model enjoys an accidental $U(1)$ symmetry
wherein one rephases $\chi$ through an arbitrary phase.
The scalar fields that compose $\chi$ are assumed
not to have any vacuum expectation value (VEV),
even if one of them---depending on $Y$ and $J$---may happen to be
electrically neutral.
There is in the scalar potential (SP) a renormalizable coupling
\be
\lambda_4\, \sum_{a=1}^3 \left( H^\dagger\, \frac{\tau_a}{2}\, H \right)
\left[ \chi^\dagger\, T_a^{(J)}\, \chi \right]
\label{coup}
\ee
of $\chi$ to the Higgs doublet $H$ of the SM.
In Eq.~\eqref{coup},
\begin{itemize}
\item $\lambda_4$ is a dimensionless coefficient,
\item the $\tau_a$ are the Pauli matrices,
\item one conceives of $\chi$ as a column vector of $n$ scalar fields,
\item the $T_a^{(J)}$ are the $n \times n$ matrices
  that represent $su(2)$ in the $J$-isospin representation.
\end{itemize}
The coupling~\eqref{coup} generates,
upon the neutral component of $H$ acquiring VEV $v$,
a squared-mass difference $\Delta m^2 \propto v^2$
between any two components of $\chi$ whose third component of isospin
differs by one unit.

This NP model was firstly (to our knowledge)
considered thirty years ago~\cite{li}
as a paradigm for potentially large oblique parameters (OPs).
It has later been studied as a model
for ``minimal'' dark matter~\cite{italians} and,
more recently~\cite{chinese},
as an explanation for the unexpectedly high value of the $W^\pm$ mass
measured by the CDF-II Collaboration.
Twelve years ago,
Logan and her collaborators~\cite{logan}
have shown that $n$ cannot exceed eight,
lest perturbative unitarity in the scattering of two scalars of $\chi$
to two $SU(2)$ gauge bosons be violated;
they also derived mixed constraints on $J$ and $Y$.
Logan's work was revived and expanded very recently~\cite{milagre}.
In another recent paper~\cite{kannikerecent},
the specific case of the addition of an $SU(2)$ scalar quadruplet
to the SM has been considered;
the hypercharge of that quadruplet
has been restricted to the values $1/2$ or $3/2$,\footnote{Other recent papers
that consider scalar quadruplets with those specific hypercharges
are Refs.~\cite{roma1,roma2,jurc}. They also consider models
with additional scalar triplets and five-plets,
always with specific hypercharges.}
because in those two cases additional quartic couplings---beyond the
one of Eq.~\eqref{coup}---of the types
$\chi H H H$ and/or $\chi \chi \chi H$ may be present.
(The accidental $U(1)$ symmetry then does not exist,
because $Y$ has a well-defined value.)
The case studied in Ref.~\cite{kannikerecent} is
on the one hand more restricted than the one in this paper,
because $\chi$ has fixed $J=3/2$,
but on the other hand it is more complicated,
because additional quartic terms are allowed in the SP.

In this paper we want to constrain the modulus of the coefficient $\lambda_4$
of the term~\eqref{coup} of the SP;
in so doing,
we place an upper bound on $\Delta m^2$.
We do this by considering both the unitarity (UNI)
and the bounded-from-below (BFB) 
conditions on the quartic part of the SP.
Remarkably,
the upper bound on $\Delta m^2$
results from both the UNI \emph{and the BFB} conditions,
and not just from the former ones.
We firstly show this fact,
in a simplified version of the SP,
in Section~\ref{without};
later on,
in Section~\ref{with},
we consider the full SP.
Section~\ref{ops} contains the confrontation of our NP model
with the OPs that it generates;
we investigate whether the phenomenological OPs constrain $\Delta m^2$
more or less than the UNI/BFB conditions.
Section~\ref{conclusions} contains our conclusions.
Appendix~\ref{uniconditions} explicitly lists the UNI conditions
for all the values of $n$ through eight.

\section{Potential without terms four-linear on $\chi$}
\label{without}

In our model of NP
there is the SM scalar doublet $H$ with hypercharge $1/2$
and an $SU(2)$ scalar multiplet $\chi$ with weak isospin $J$,
which is a positive number,
either integer or half-integer.
The multiplet $\chi$ has
\be
\label{J}
n = 2 J + 1
\ee
components $\chi_I$
($I = J,\ J-1,\ J-2,\ \ldots,\ 1 - J,\ -J$).
Its hypercharge $Y$ remains un-specified,
\textit{i.e.}\ arbitrary.
Together with the charge-conjugate multiplets $\tilde H$
and $\tilde \chi$,
we have the four multiplets
\be
H = \left( \begin{array}{c} a \\ b \end{array} \right), \quad
\tilde H = \left( \begin{array}{c} b^* \\ - a^* \end{array} \right), \quad
\chi = \left( \begin{array}{c}
  \chi_J \\ \chi_{J-1} \\ \chi_{J-2} \\ \vdots \\ \chi_{1-J} \\ \chi_{-J}
\end{array} \right), \quad
\tilde \chi = \left( \begin{array}{c}
  \chi_{-J}^\ast \\ - \chi_{1-J}^\ast \\ \vdots \\
  \left( - 1 \right)^{2J} \chi_{J-2}^\ast \\
  - \left( - 1 \right)^{2J} \chi_{J-1}^\ast \\
  \left( - 1 \right)^{2J} \chi_J^\ast
\end{array} \right).
\ee
Here,
$a$,
$b$,
and the $\chi_I$ are complex Klein--Gordon fields.
Their third components of isospin are
\be
\label{T3}
a: \frac{1}{2}, \quad
b: - \frac{1}{2}, \quad
\chi_I: I, \quad
\chi_I^\ast: -I.
\ee

When one multiplies $\chi$ by $\tilde \chi$ one obtains,
among other $SU(2)$ representations,
the singlet
\be
\label{ffff2}
F_2 \equiv \left( \chi \otimes \tilde \chi \right)_\mathbf{1}
= \sum_{I=-J}^J \left| \chi_I \right|^2
\ee
and the triplet
\be
\label{uopc}
\left( \chi \otimes \tilde \chi \right)_\mathbf{3}
= \left( \begin{array}{c}
  \displaystyle{- \sum_{I=1-J}^J
    \sqrt{\frac{J^2 - I^2 + J + I}{2}}\
    \chi_I\, \chi_{I-1}^\ast}
  \\*[2mm]
  \displaystyle{\sum_{I=-J}^J I \left| \chi_I \right|^2}
  \\*[2mm]
  \displaystyle{\sum_{I=1-J}^J
    \sqrt{\frac{J^2 - I^2 + J + I}{2}}\
  \chi_I^\ast\, \chi_{I-1}}
\end{array} \right).
\ee
Applying the general Eq.~\eqref{uopc} to the specific case of $H$
(\textit{i.e.},
using $J=1/2$,
$\chi_{1/2} = a$,
and $\chi_{-1/2} = b$),
we obtain
\be
\left( H \otimes \tilde H \right)_\mathbf{3} = \left( \begin{array}{c}
  \displaystyle{- \frac{a b^\ast}{\sqrt{2}}}
  \\*[2mm]
  \displaystyle{\frac{\left| a \right|^2 - \left| b \right|^2}{2}}
  \\*[2mm]
  \displaystyle{\frac{a^\ast b}{\sqrt{2}}}
\end{array} \right).
\ee

The SP $V$ has a quadratic part $V_2$ and a quartic part $V_4$:
\be
\label{v5}
V = V_2 + V_4.
\ee
Obviously,
\be
\label{v}
V_2 = - \mu_1^2 F_1 + \mu_2^2 F_2,
\ee
where
\be
\label{ffff1}
F_1 \equiv \left( H \otimes \tilde H \right)_\mathbf{1}
= \left| a \right|^2 + \left| b \right|^2
\ee
and $F_2$ is defined in Eq.~\eqref{ffff2}.
We assume both coefficients $\mu_1^2$ and $\mu_2^2$ to be positive,
so that $H$ has VEV $\left\langle 0 \left| b \right| 0 \right\rangle = v$
but $\chi$ does not have VEV.

The quartic part of the SP contains
\begin{itemize}
\item the term
  $\left[ \left( H \otimes \tilde H \right)_\mathbf{1} \right]^2$,
  with coefficient $\displaystyle{\frac{\lambda_1}{2}}$;
\item the term
  $\left( H \otimes \tilde H \right)_\mathbf{1}\,
  \left( \chi \otimes \tilde \chi \right)_{\mathbf{1}}$,
  with coefficient $\lambda_3$;
\item the term
  $\left[ \left( H \otimes \tilde H \right)_\mathbf{3} \otimes
  \left( \chi \otimes \tilde \chi \right)_{\mathbf{3}}
  \right]_{\mathbf{1}}$,
  with coefficient $\lambda_4$;
\item various terms that are four-linear in the components of $\chi$.
  We keep those terms un-specified in this section.
\end{itemize}
Thus,
\be
V_4 =
\frac{\lambda_1}{2}\, F_1^2
+ \lambda_3 F_1 F_2
+ \lambda_4 F_4
+ \mbox{terms\ four-linear\ in\ the}\ \chi_I,
\label{poiu}
\ee
where
\be
F_4 \equiv \frac{\left| a \right|^2 - \left| b \right|^2}{2}\,
\sum_{I=-J}^J I \left| \chi_I \right|^2
+ \frac{z + z^\ast}{2}.
\label{ffff4}
\ee
We have defined
\be
z \equiv
a b^\ast \sum_{I=1-J}^J \chi_I^\ast\, \chi_{I-1}
\sqrt{J^2 - I^2 + J + I}.
\label{z}
\ee

From Eqs.~\eqref{v5},
\eqref{v},
\eqref{poiu},
and~\eqref{ffff4} the mass-squared of the scalar $\chi_I$ is
\be
m_I^2 = \mu_2^2 + \left( \lambda_3 - \frac{\lambda_4}{2}\, I \right)
\left| v \right|^2.
\ee
This implies that the difference between the masses-squared
of $\chi_I$ and $\chi_{I+1}$ is
\be
\Delta m^2 = \frac{\left| \lambda_4 v^2 \right|}{2},
\label{deltam}
\ee
which is $I$-independent.
An upper bound on $\left| \lambda_4 \right|$
is therefore equivalent to an upper bound on $\Delta m^2$.

The VEV of $V$ is
\be
\left\langle 0 \left| V \right| 0 \right\rangle
= - \mu_1^2 v^2 + \frac{\lambda_1}{2}\, v^4.
\ee
Therefore,
$\mu_1^2 = \lambda_1 v^2$.
The mass-squared of the Higgs particle is $m_H^2 = 2 \lambda_1 v^2$.
Since experimentally $m_H \approx$ 125\,GeV and $v \approx$ 174\,GeV,
one has
\be
\lambda_1 \approx 0.258.
\label{lambda1}
\ee
From now on we shall assume Eq.~\eqref{lambda1} to hold.
Contrary to $\lambda_1$,
the couplings $\lambda_3$ and $\lambda_4$ are free,
but they are constrained by both the UNI and BFB conditions.
We next derive those constraints.

\subsection{Unitarity (UNI) conditions}

In~\cite{logan},
and more recently again in~\cite{milagre},
the scattering of two scalars belonging to $\chi$
to two gauge bosons of either gauge group $SU(2)$ or $U(1)$ has been considered;
therefrom upper bounds on both the isospin $J$
and the hypercharge $Y$ of $\chi$ have been derived.
Here we consider the scattering of a pair of scalars of $\chi$
to another pair of scalars,
both pairs having,
of course,
the same $I$ (third component of isospin) and $Y$.
While in~\cite{logan,milagre} the scattering involves
two cubic gauge couplings and
the interchange of a virtual particle either in the $s$,
$t$,
or $u$ channel,
here the scattering involves no interchange of any virtual particle,
rather it takes place directly through a \emph{quartic} coupling in the SP.

Firstly suppose that $J$ is half-integer.
\begin{itemize}
\item We consider the scattering of the two
  two-field states with hypercharge $Y + 1/2$
  and null third component of isospin,
  \textit{viz.}\ of $\chi_{-1/2} a$ and $\chi_{1/2} b$.
  Their scattering matrix is
  \be
  \left( \begin{array}{cc}
    \lambda_3 - \lambda_4 / 4 & \left( 2 J + 1 \right) \lambda_4 / 4 \\
    \left( 2 J + 1 \right) \lambda_4 / 4 & \lambda_3 - \lambda_4 / 4
  \end{array} \right).
  \ee
  The eigenvalues of this matrix are
  \be
  \label{ev1}
  \lambda_3 + \frac{J \lambda_4}{2},
  \quad
  \lambda_3 - \frac{\left( J + 1 \right) \lambda_4}{2}.
  \ee
\item We next consider the scattering of the states
  with hypercharge $Y - 1/2$ and null third component of isospin,
  \textit{viz.}\ of $\chi_{-1/2} b^\ast$ and $\chi_{1/2} a^\ast$.
  Their scattering matrix is
  \be
  \left( \begin{array}{cc}
    \lambda_3 + \lambda_4 / 4 & \left( 2 J + 1 \right) \lambda_4 / 4 \\
    \left( 2 J + 1 \right) \lambda_4 / 4 & \lambda_3 + \lambda_4 / 4
  \end{array} \right).
  \ee
  The eigenvalues of this matrix are
  \be
  \label{ev2}
  \lambda_3 + \frac{\left( J + 1 \right) \lambda_4}{2},
  \quad
  \lambda_3 - \frac{J \lambda_4}{2}.
  \ee
\end{itemize}

Let secondly suppose that $J$ is an integer instead.
\begin{itemize}
\item We consider the scattering of the two two-particle states
  with hypercharge $Y + 1/2$ and third component of isospin $1/2$,
  \textit{viz.}\ of $\chi_0 a$ and $\chi_1 b$.
  Their scattering matrix is
  \be
  \left( \begin{array}{cc}
    \lambda_3 & \left. \sqrt{J \left( J + 1 \right)}\, \lambda_4 \right/ 2 \\
    \left. \sqrt{J \left( J + 1 \right)}\, \lambda_4 \right/ 2 &
    \lambda_3 - \lambda_4 / 2
  \end{array} \right).
  \ee
  The eigenvalues of this matrix are the ones in Eq.~\eqref{ev1}.
\item We next consider the scattering of the states
  with hypercharge $Y - 1/2$ and third component of isospin $1/2$,
  \textit{viz.}\ of the states $ \chi_1 a^\ast$ and $\chi_0 b^\ast$.
  Their scattering matrix is
  \be
  \left( \begin{array}{cc}
    \lambda_3 + \lambda_4 / 2 &
    \left. \sqrt{J \left( J + 1 \right)}\, \lambda_4 \right/ 2 \\
    \left. \sqrt{J \left( J + 1 \right)}\, \lambda_4 \right/ 2 & \lambda_3
  \end{array} \right).
  \ee
  The eigenvalues of this matrix are in Eq.~\eqref{ev2}.
\end{itemize}

Thus,
the eigenvalues of the scattering matrices are the same,
no matter whether $J$ is integer or half-integer.

We now impose the conditions that the moduli of all the eigenvalues
in Eqs.~\eqref{ev1} and~\eqref{ev2} should be smaller than
\be
M = 8 \pi.
\label{M}
\ee
We obtain
\bs
\label{uniarb}
\ba
\left| \lambda_3 \right| + \frac{J}{2} \left| \lambda_4 \right| &<& M,
\label{uniarb1}
\\
\left| \lambda_3 \right| + \frac{J+1}{2} \left| \lambda_4 \right| &<& M.
\label{uniarb2}
\ea
\es
Condition~\eqref{uniarb2} is of course stronger
than condition~\eqref{uniarb1},
therefore one may neglect the latter.

The dispersion of the $2 + n$ states that have zero
third component of isospin and zero hypercharge,
\textit{viz.}\ of the states\footnote{In this explicit computation
we assume $J$ to be an integer.
The final result,
\textit{viz}.~Eqs.~\eqref{uniarbprime},
is also valid for half-integer $J$.}
\be
\left| a \right|^2, \left| b \right|^2,
\left| \chi_J \right|^2, \left| \chi_{-J} \right|^2,
\left| \chi_{J-1} \right|^2, \left| \chi_{1-J} \right|^2,
\ldots,
\left| \chi_1 \right|^2, \left| \chi_{-1} \right|^2,
\left| \chi_0 \right|^2
\ee
produces the scattering matrix
\be
\mathcal{S} = \left( \begin{array}{cccccc}
  \mathcal{A} & \mathcal{B}_J & \mathcal{B}_{J-1} & \cdots &
  \mathcal{B}_1 & \mathcal{C} \\
  \mathcal{B}_J & 0_{2 \times 2} & 0_{2 \times 2} & \cdots &
  0_{2 \times 2} & 0_{2 \times 1} \\
  \mathcal{B}_{J-1} & 0_{2 \times 2} & 0_{2 \times 2} & \cdots &
  0_{2 \times 2} & 0_{2 \times 1} \\
  \vdots & \vdots & \vdots & \ddots & \vdots & \vdots \\ 
  \mathcal{B}_1 & 0_{2 \times 2} & 0_{2 \times 2} & \cdots &
  0_{2 \times 2} & 0_{2 \times 1} \\
  \mathcal{C}^T & 0_{1 \times 2} & 0_{1 \times 2} & \cdots &
  0_{1 \times 2} & 0
\end{array} \right),
\label{jbogfof}
\ee
where
\be
\mathcal{A} = \left( \begin{array}{cc}
  2 \lambda_1 & \lambda_1 \\ \lambda_1 & \lambda_1
\end{array} \right),
\quad
\mathcal{B}_I = \left( \begin{array}{cc}
  \lambda_3 + I \lambda_4 / 2 & \lambda_3 - I \lambda_4 / 2 \\
  \lambda_3 - I \lambda_4 / 2 & \lambda_3 + I \lambda_4 / 2
\end{array} \right),
\quad
\mathcal{C} = \left( \begin{array}{c}
  \lambda_3 \\ \lambda_3
\end{array} \right),
\ee
and $0_{m \times m^\prime}$ denotes the $m \times m^\prime$ matrix
that has all its matrix elements equal to zero.
The matrix $\mathcal{S}$ is equivalent to
\be
\left( \begin{array}{cccccc}
  X \mathcal{A} X^T & X \mathcal{B}_J X^T & X \mathcal{B}_{J-1} X^T & \cdots &
  X \mathcal{B}_1 X^T & X \mathcal{C} \\
  X \mathcal{B}_J X^T & 0_{2 \times 2} & 0_{2 \times 2} & \cdots &
  0_{2 \times 2} & 0_{2 \times 1} \\
  X \mathcal{B}_{J-1} X^T & 0_{2 \times 2} & 0_{2 \times 2} & \cdots &
  0_{2 \times 2} & 0_{2 \times 1} \\
  \vdots & \vdots & \vdots & \ddots & \vdots & \vdots \\ 
  X \mathcal{B}_1 X^T & 0_{2 \times 2} & 0_{2 \times 2} & \cdots &
  0_{2 \times 2} & 0_{2 \times 1} \\
  \mathcal{C}^T X^T & 0_{1 \times 2} & 0_{1 \times 2} & \cdots &
  0_{1 \times 2} & 0
\end{array} \right),
\ee
where
\be
X = \frac{1}{\sqrt{2}} \left( \begin{array}{cc}
  1 & 1 \\ - 1 & 1
\end{array} \right)
\ee
and consequently
\be
X \mathcal{A} X^T = \left( \begin{array}{cc} 3 \lambda_1 & 0 \\ 0 & \lambda_1
  \end{array} \right),
\quad
X \mathcal{B}_I X^T = \left( \begin{array}{cc} 2 \lambda_3 & 0 \\
  0 & I \lambda_4
  \end{array} \right),
\quad
X \mathcal{C} X = \left( \begin{array}{c} \sqrt{2} \lambda_3 \\ 0
\end{array} \right).
\ee
Thus,
the matrix $S$ is equivalent to the direct sum of the two matrices
$\mathcal{S}_+$ and $\mathcal{S}_-$,
where
\bs
\ba
\mathcal{S}_+ &=& \left( \begin{array}{cccccc}
  3 \lambda_1 & 2 \lambda_3 & 2 \lambda_3 &
  \cdots & 2 \lambda_3 & \sqrt{2} \lambda_3 \\
  2 \lambda_3 & 0 & 0 & \cdots & 0 & 0 \\
  2 \lambda_3 & 0 & 0 & \cdots & 0 & 0 \\
  \vdots & \vdots & \vdots & \ddots & \vdots & \vdots \\
  2 \lambda_3 & 0 & 0 & \cdots & 0 & 0 \\
  \sqrt{2} \lambda_3 & 0 & 0 & \cdots & 0 & 0
\end{array} \right),
\\
\mathcal{S}_- &=& \left( \begin{array}{ccccc}
  \lambda_1 & J \lambda_4 & \left( J - 1 \right) \lambda_4 &
  \cdots & \lambda_4 \\
  J \lambda_4 & 0 & 0 & \cdots & 0 \\
  \left( J - 1 \right) \lambda_4 & 0 & 0 & \cdots & 0 \\
  \vdots & \vdots & \vdots & \ddots & \vdots \\
  \lambda_4 & 0 & 0 & \cdots & 0 \\
\end{array} \right).
\ea
\es
Computing the eigenvalues of these two matrices
and setting their moduli to be smaller than $M$,
we find
\bs
\ba
3 \left| \lambda_1 \right|
+ \sqrt{9 \lambda_1^2 + 4 \left[ \sum_{I=1}^J \left( 4 \lambda_3^2 \right)
+ 2 \lambda_3^2 \right]} &<& 2 M,
\\
\left| \lambda_1 \right|
+ \sqrt{ \lambda_1^2 + 4 \lambda_4^2 \sum_{I=1}^J I^2} &<& 2 M.
\ea
\es
Performing the sums over $I$,
we obtain
\bs
\label{uniarbprime}
\ba
3 \left| \lambda_1 \right|
+ \sqrt{9 \lambda_1^2 + 8 \left( 2 J + 1 \right) \lambda_3^2} &<& 2 M,
\label{uniarb4}
\\
\left| \lambda_1 \right|
+ \sqrt{\lambda_1^2
  + \frac{2}{3}\, J \left( J + 1 \right) \left( 2 J + 1 \right) \lambda_4^2}
&<& 2 M.
\label{uniarb3}
\ea
\es

The potential $V_4$ produces many other scatterings of two-particle states,
but they all lead to UNI conditions that either repeat Eq.~\eqref{uniarb2},
or repeat Eq.~\eqref{uniarb3},
or repeat Eq.~\eqref{uniarb4},
or are weaker than one of them.

\subsection{Bounded-from-below (BFB) conditions}

In order to evaluate the BFB conditions on $V_4$,
it is handy to use the gauge wherein $b = 0$.
We use
\bs
\ba
\left| \sum_{I=-J}^J I \left| \chi_I \right|^2 \right|
&=& \left| J \left| \chi_J \right|^2 
+ \left( J - 1 \right) \left| \chi_{J-1} \right|^2 
+ \cdots - J \left| \chi_{-J} \right|^2 \right|
\\
&\le& J F_2
\ea
\es
to write,
in that gauge,
\be
V_4 \ge
\frac{\lambda_1}{2} \left| a \right|^4
+ \left( \lambda_3 - \left. J \left| \lambda_4 \right| \right/ 2 \right)
\left| a \right|^2 F_2
+ \mbox{terms\ four-linear\ in\ the}\ \chi_I.
\label{uq}
\ee
Since both $\left| a \right|^2 \ge 0$ and $F_2 \ge 0$,
the conditions for $\left( \lambda_1 / 2 \right) \left| a \right|^4
+ \left( \lambda_3 - \left. J \left| \lambda_4 \right| \right/ 2 \right)
\left| a \right|^2 F_2$ to be non-negative,
whatever the (non-negative) values of $\left| a \right|^2$ and $F_2$,
are~\cite{kannike}
\bs
\label{fbfbbfb}
\ba
\lambda_1 &\ge& 0, \\
\lambda_3 &\ge& 0, \label{www3} \\
\left| \lambda_4 \right| &\le& \frac{2 \lambda_3}{J}. \label{www4}
\ea
\es
Conditions~\eqref{fbfbbfb} are \emph{necessary and sufficient}
for $V_4$ to be non-negative for arbitrary values of the fields $a$
and $\chi_I$.
Since a gauge with $b=0$ can always be obtained,
those conditions also hold
for $V_4$ with arbitrary values of $a$,
$b$,
and the $\chi_I$.

\subsection{Results}

Transforming all the six relevant UNI and BFB
conditions into strict equalities,
we have
\bs
\ba
\lambda_3 + \frac{J + 1}{2} \left| \lambda_4 \right| &=& M,
\label{primeira} \\
\lambda_1 + \sqrt{\lambda_1^2 + \frac{2 J \left( 2 J^2 + 3 J + 1 \right)
  \lambda_4^2}{3}} &=& 2 M,
\label{segunda} \\
3 \lambda_1 + \sqrt{9 \lambda_1^2 + 8 \left( 2 J + 1 \right)
  \lambda_3^2} &=& 2 M,
\label{terceira} \\
\left| \lambda_4 \right| &=& \frac{2}{J}\, \lambda_3,
\label{quarta}
\ea
\es
where we have taken into account that both $\lambda_1$ and $\lambda_3$
are non-negative,
\textit{cf.}\ Eqs.~\eqref{lambda1} and~\eqref{www3},
respectively.
Equation~\eqref{segunda} gives solution I:
\be
\left| \lambda_4 \right| = \sqrt{\frac{6 M
    \left( M - \lambda_1 \right)}{J \left( 2 J^2 + 3 J + 1 \right)}}.
\label{um}
\ee
Equations~\eqref{primeira} and~\eqref{terceira} together produce solution~II:
\bs
\label{dois}
\ba
\left| \lambda_4 \right| &=& \displaystyle{\frac{2 M}{J + 1}
- \frac{\sqrt{2 M \left( M - 3 \lambda_1 \right)}}{\left(
  J + 1 \right) \sqrt{2 J + 1}},}
\\
\lambda_3 &=& \displaystyle{\sqrt{\frac{M \left( M - 3 \lambda_1 \right)}{2
      \left( 2 J + 1 \right)}}.}
\label{cio1}
\ea
\es
Equations~\eqref{primeira} and~\eqref{quarta} together lead to solution~III:
\bs
\label{tres}
\ba
\left| \lambda_4 \right| &=& \displaystyle{\frac{2 M}{2 J + 1},}
\label{IIIlambda4}
\\
\lambda_3 &=& \displaystyle{\frac{J}{2 J + 1}\, M.}
\ea
\es
Equations~\eqref{terceira} and~\eqref{quarta} together give solution~IV:
\be
\label{quatro}
\left| \lambda_4 \right| = \displaystyle{\frac{\sqrt{2 M
    \left(M - 3 \lambda_1 \right)}}{J \sqrt{2 J + 1}}}
\ee
and Eq.~\eqref{cio1},
which is the solution of Eq.~\eqref{terceira}.

\begin{figure}[ht]
\begin{center}
\includegraphics[width=1.0\textwidth]{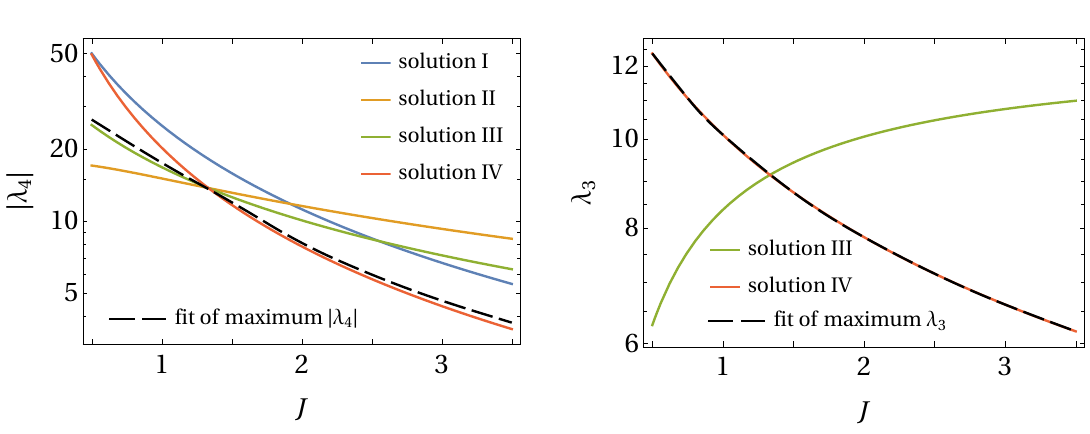}
\end{center}
\caption{The solutions~\eqref{um}--\eqref{quatro}
  for $\left| \lambda_4 \right|$ (left panel)
  and $\lambda_3$ (right panel)
  \textit{versus} $J$.
  We have used $M = 8 \pi$.
  The black dashed lines display the maximum allowed values
  of $\left| \lambda_4 \right|$ and $\lambda_3$
  described in subsection~\ref{sect:results}.}
\label{fig:lmd4_solut}
\end{figure}

Solutions~I,
II,
III,
and IV for $\left| \lambda_4 \right|$ and $\lambda_3$
are plotted in Fig.~\ref{fig:lmd4_solut}.
Notice that solutions~II,
III,
and~IV coincide when
\be
\label{nmax}
2 J + 1 = \frac{2 M J^2}{M - 3 \lambda_1},
\ee
\textit{i.e.}\ when $J \approx 1.37$.
When $J$ is larger than this value,
\textit{i.e.}\ when $\chi$ is either a quadruplet or a larger multiplet
of $SU(2)$,
then solution~IV---which arises from both
the UNI condition~\eqref{uniarb4}
and the BFB condition~\eqref{www4}---gives
the strongest upper bound on both $\lambda_3$ and $\left| \lambda_4 \right|$.
Thus,
both $\lambda_3$ and $\left| \lambda_4 \right|$ are bounded from above
by the interplay of a UNI condition and a BFB condition,
for most possible values of $J$.

We remind the reader that,
according to Eq.~\eqref{www3},
the minimum value of $\lambda_3$ is 0.

\subsection{Renormalization-group equations}

In a renormalizable field theory,
the values of the dimensionless coupling constants
evolve with the energy scale $\mu$ at which they are measured.
That evolution is governed by differential equations named
renormalization-group equations (RGEs)\footnote{Here we only consider
the one-loop-level RGEs.}:
\be
16 \pi^2 \mu\, \frac{\mathrm{d} g}{\mathrm{d} \mu} = \beta_g,
\ee
where $g$ denotes a generic dimensionless coupling
and $\beta_g$ is a function of,
in general,
all the dimensionless couplings in the theory.
Formulas for the functions $\beta_g$ in a general gauge theory
have been presented long ago by Cheng, Eichten, and Li~\cite{chengli}.
In the case at hand there is a gauge theory
with gauge group $SU(3) \times SU(2) \times U(1)$
and gauge coupling constants $g_3$ for $SU(3)$,
$g_2$ for $SU(2)$,
and $g_1$ for $U(1)$;\footnote{We use here the normalization
for $g_1$ usual in the $SU(5)$,
$SO(10)$,
and $E_6$ Grand Unified Theories.
Still,
we keep for the hypercharges of the multiplets the usual normalization
given by $Q = I + Y$,
where $Q$ is a field's electric charge,
$I$ is the third component of isospin,
and $Y$ is the hypercharge.}
in that gauge theory there is an $SU(2)$ doublet with hyperchage $1/2$
and an $SU(2)$ multiplet with isospin $J$ and hypercharge $Y$.
We consider in this subsection the SP
\be
V = \sum_{k=1}^2 \left( \mu_k^2 F_k + \frac{\lambda_k}{2}\, F_k^2 \right)
+ \lambda_3 F_1 F_2 + \lambda_4 F_4,
\ee
\textit{i.e.}\ we take into account the presence in $V$
of a quartic term proportional to $F_2^2$,
but we discard all the other terms four-linear in the $\chi_I$
(they are studied in some detail in section~\ref{with}).\footnote{In
the numerical part of this section,
though,
we shall fix $\lambda_2 = 0$ for the sake of simplicity.}
The dimensionless couplings that we take into account are,
thus,
$g_1$,
$g_2$,
$g_3$,
$\lambda_1$,
$\lambda_2$,
$\lambda_3$,
and $\lambda_4$.
Additionally,
in the full theory there are fermions
with Yukawa couplings to the scalar doublet,
and one (and only one) of those Yukawa couplings,
\textit{viz.}~$y_t$---the Yukawa coupling of $H$ to the top quark---is
rather large and therefore has a strong influence on the RGEs;
so,
there is one further dimensionless coupling $y_t$ that we take into account.
In order to derive the RGEs for these eight coupling constants
we have used a feature of the software {\tt SARAH}~\cite{sarah}.
(We point out that that software only tolerates $SU(2)$ multiplets
with isospin up to 3,
so we had to edit it and make a modification in order to derive
the RGEs for the case $J = 7/2$.
We moreover point out that the running time for that software
increases \emph{exponentially} with the size of the $SU(2)$ multiplets.)
We have obtained
\bs
\label{g}
\ba
\beta_{g_1} &=& \left( \frac{41}{10} + \frac{4}{5}\, Y^2 \right) g_1^3,
\\
\beta_{g_2} &=& \left[ - \frac{19}{6}
+ \frac{J \left( J + 1 \right) \left( 2 J + 1 \right)}{9} \right] g_2^3,
\\
\beta_{g_3} &=& - 7 g_3^3,
\ea
\es
\bs
\label{l}
\ba
\beta_{\lambda_1} &=&
\frac{27}{100}\, g_1^4
+ \frac{9}{10}\, g_1^2 g_2^2
+ \frac{9}{4}\, g_2^4
+ 12 y_t^2 \lambda_1
+ 12 \lambda_1^2
+ 2 \left( 2 J + 1 \right) \lambda_3^2
+ \frac{J \left( J + 1 \right) \left( 2 J + 1 \right)}{6}\, \lambda_4^2
\no & &
- \left( \frac{9}{5}\, g_1^2 + 9 g_2^2 \right) \lambda_1 - 12 y_t^4,
\\
\beta_{\lambda_2} &=&
\frac{108}{25} \left( Y g_1 \right)^4
+ \frac{72}{5} \left( Y g_1 \right)^2 \left( J g_2 \right)^2
+ 6 J^2 \left( 2 J^2 + 1 \right) g_2^4
+ 2 \left( 2 J + 5 \right) \lambda_2^2
+ 4 \lambda_3^2
+ J^2 \lambda_4^2
\no & &
- \left[ \frac{36}{5} \left( Y g_1 \right)^2
+ 12 J \left( J + 1 \right) g_2^2 \right] \lambda_2,
\\
\beta_{\lambda_3} &=&
\frac{27}{25} \left( Y g_1^2 \right)^2
+ 3 J \left( J + 1 \right) g_2^4
+ 6 y_t^2 \lambda_3
+ 6 \lambda_1 \lambda_3
+ 4 \left( J + 1 \right) \lambda_2 \lambda_3
+ 4 \lambda_3^2
+ J \left( J + 1 \right) \lambda_4^2
\no & &
- \left( \frac{9}{10} + \frac{18}{5}\, Y^2 \right) g_1^2 \lambda_3
- \left[ \frac{9}{2} + 6 J \left( J + 1 \right) \right] g_2^2 \lambda_3,
\\
\beta_{\lambda_4} &=&
\frac{36}{5}\, Y g_1^2 g_2^2
+ 6 y_t^2 \lambda_4
+ 2 \lambda_1 \lambda_4
+ 8 \lambda_3 \lambda_4
\no & &
- \left( \frac{9}{10} + \frac{18}{5}\, Y^2 \right) g_1^2 \lambda_4
- \left[ \frac{9}{2} + 6 J \left( J + 1 \right) \right] g_2^2 \lambda_4,
\ea
\es
\be
\label{y}
\beta_{y_t} = \left( \frac{9}{2}\, y_t^2
- \frac{17}{20}\, g_1^2 - \frac{9}{4}\, g_2^2 - 8 g_3^2 \right) y_t.
\ee

\begin{figure}[ht]
\begin{center}
\includegraphics[width=1.0\textwidth]{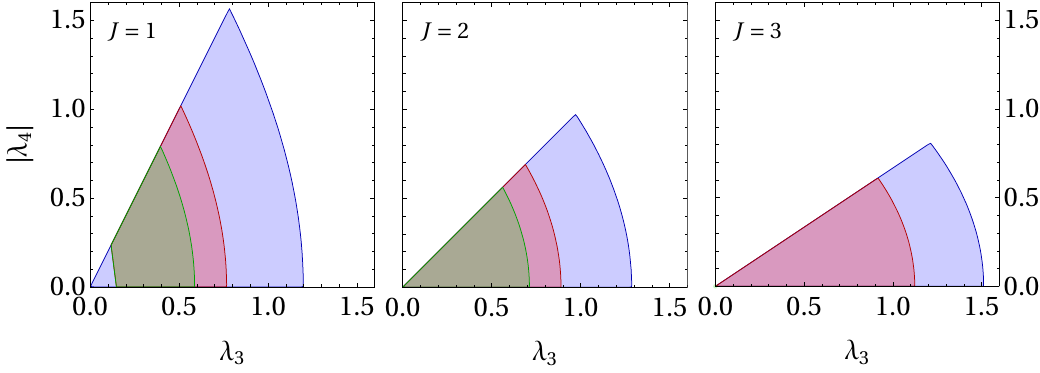}
\end{center}
\caption{Regions of stability for $J = 1$ (left panel),
  $J =2$ (middle panel),
  and $J = 3$ (right panel)
  and for energy ranges from $\mu = m_t$ to $10^9$~GeV (blue regions),
  $10^{14}$~GeV (red regions),
  and $10^{19}$~GeV (green regions).}
\label{fig:regions_lmd4-lmd3}
\end{figure}

We have used a numerical code to solve
the differential equations~\eqref{g}--\eqref{y}
starting at the scale $\mu = m_t = 173.1$\,GeV 
and letting $\mu$ evolve up to the scale
$\mu_\mathrm{Planck} = 10^{19}$\,GeV.
We have slightly simplified those equations by setting $\lambda_2 = 0$
at all $\mu$ values---even though in general $\beta_{\lambda_2}$
is nonzero and therefore a nonzero $\lambda_2$ will be generated
even if one starts with $\lambda_2 = 0$---and
by assuming the hypercharge $Y$ of the additional multiplet
to be zero too.
At $\mu = m_t$ we have fixed~\cite{huang}
\bs
\label{fixed_couplings}
\ba
&& g_1 = \sqrt{\frac{5}{3}} \times 0.358545, \quad
g_2 = 0.64765, \quad
g_3 = 1.1618,
\\
&& \lambda_1 = 0.258, \quad
y_t = \frac{\sqrt{2}\times 161.98\,\mathrm{GeV}}{246\,\mathrm{GeV}},
\ea
\es
and we have let $\lambda_3$ and $\lambda_4$ vary freely
while obeying the UNI and BFB conditions.
Moreover,
we have enforced the UNI and BFB conditions
at every intermediate scale $\mu$;
this indirectly constrains the initial $\lambda_3$ and $\lambda_4$ because,
if they are too large,
then at some intermediate $\mu < \mu_\mathrm{GUT}$
either the UNI or the BFB conditions will be broken.
\begin{figure}[ht]
\begin{center}
\includegraphics[width=1.0\textwidth]{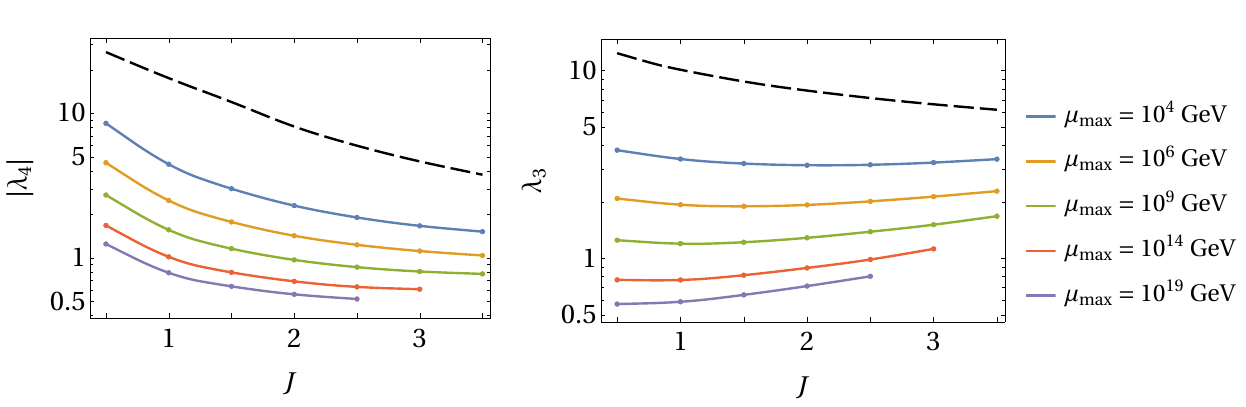}
\end{center}
\caption{The maximum allowed values of $\left| \lambda_4 \right|$ (left panel)
  and $\lambda_3$ (right panel)
  \textit{versus} $J$ for different cut-off scales $\mu_\mathrm{max}$.
  The black dashed lines are the same as in Figure~\ref{fig:lmd4_solut};
  they display the maximum values at the electroweak scale.}
\label{fig:RGEs_lmd4-lmd3}
\end{figure} 
In Figure~\ref{fig:RGEs_lmd4-lmd3} one observes the result of this labor
in the form of upper bounds on $\lambda_3$ and $\left| \lambda_4 \right|$
at the scale $\mu = m_t$,
depending on the scale $\mu_\mathrm{maximum}$ at which either the UNI
or the BFB conditions start being broken.
As expected,
if one demands that the UNI and BFB conditions are respected
for a longer $\mu$ range,
then one obtains ever stricter upper bounds on the initial values
of $\lambda_3$ and $\lambda_4$.
The same upper bounds may be also observed,
now in a correlated fashion,
for three values of the $\mu_\mathrm{max}$,
in Figure~\ref{fig:regions_lmd4-lmd3}.

The {\tt SARAH} model files and output files,
and the expressions of the RGEs for \emph{all} quartic couplings
(including the ones mentioned in section~\ref{with}),
both in pdf and {\tt Mathematica} notebook files,
are available at
\href{https://github.com/jurciukonis/RGEs-for-multiplets}{{\tt
    https://github.com/jurciukonis/RGEs-for-multiplets}}.

\section{Full potential}
\label{with}

The product $\chi \otimes \chi$ of two \emph{identical} multiplets
of $SU(2)$ only has a \emph{symmetric} component---the anti-symmetric component
vanishes because the two multiplets are equal---which consists
of\footnote{The ceiling function
$\mathrm{ceil} \left( n, 2 \right)$ maps $n$
into the smallest multiple of 2 larger than or equal to $n$.}
\be
t = \frac{1}{2}\, \mathrm{ceil} \left( n, 2 \right)
\ee
multiplets of $SU(2)$
%
\be
\left( \chi \otimes \chi \right)_\mathrm{symmetric}
= c \oplus d \oplus e \oplus \cdots \oplus q,
\ee
where
\begin{itemize}
\item $c$ is an $SU(2)$ multiplet with weak isospin $2 J$,
\item $d$ is an $SU(2)$ multiplet with weak isospin $2 J - 2$,
\item $e$ is an $SU(2)$ multiplet with weak isospin $2 J - 4$,
\end{itemize}
and so on;
lastly,
$q$ is either a triplet of $SU(2)$ if $J$ is half-integer,
or $SU(2)$-invariant if $J$ is integer.
Thus,
\be
\left( \chi \otimes \chi \right)_\mathrm{symmetric}
= \left( \begin{array}{c}
  c_{2J} \\ c_{2J-1} \\ \vdots \\ c_{-2J}
\end{array} \right) 
\oplus 
\left( \begin{array}{c}
  d_{2J-2} \\ d_{2J-3} \\ \vdots \\ d_{2-2J}
\end{array} \right)  
\oplus 
\left( \begin{array}{c}
  e_{2J-4} \\ e_{2J-5} \\ \vdots \\ e_{4-2J}
\end{array} \right)  
\oplus \cdots,
\label{symm_mult}
\ee
where the sub-indices give the third component of isospin.
The two-field states in each multiplet
in the right-hand side of Eq.~\eqref{symm_mult}
are evaluated by using Clebsch--Gordan
coefficients in the standard fashion.
Thus,
\bs
\label{ci}
\ba
c_I &=& \sum_{I^\prime = - J}^J\, \sum_{I^{\prime \prime} = - J}^J\,
\delta_{I, I^\prime + I^{\prime \prime}}
\left[ \begin{array}{ccc}
    J & J & 2 J \\
    I^\prime & I^{\prime \prime} & I
  \end{array} \right] \chi_{I^\prime} \chi_{I^{\prime \prime}},
\\
d_I &=& \sum_{I^\prime = - J}^J\, \sum_{I^{\prime \prime} = - J}^J\,
\delta_{I, I^\prime + I^{\prime \prime}}
\left[ \begin{array}{ccc}
    J & J & 2 J - 2 \\
    I^\prime & I^{\prime \prime} & I
  \end{array} \right] \chi_{I^\prime} \chi_{I^{\prime \prime}},
\\
e_I &=& \sum_{I^\prime = - J}^J\, \sum_{I^{\prime \prime} = - J}^J\,
\delta_{I, I^\prime + I^{\prime \prime}}
\left[ \begin{array}{ccc}
    J & J & 2 J - 4 \\
    I^\prime & I^{\prime \prime} & I
  \end{array} \right] \chi_{I^\prime} \chi_{I^{\prime \prime}},
\ea
\es
and so on.

The ``terms four-linear in the $\chi_I$'' in Eq.~\eqref{poiu} are
\begin{itemize}
\item a term $\displaystyle{\frac{\lambda_2}{2}\, F_2^2}$,
  where $F_2$ has been defined in Eq.~\eqref{ffff2};
\item a term $\lambda_5 F_5$,
  where
  \be
  F_5 \equiv \sum_{I=2-2J}^{2 J - 2} \left| d_I \right|^2;
  \label{ffff5}
  \ee
\item a term $\lambda_6 F_6$,
  where
  \be
  F_6 \equiv \sum_{I=4-2J}^{2 J - 4} \left| e_I \right|^2;
  \label{ffff6}
  \ee
\item and other analogous terms,
  up to $\lambda_{t+3} F_{t+3}$,
  where
  \be
  F_{t+3} \equiv \left\{ \begin{array}{lcl}
    \left| q_1 \right|^2 + \left| q_0 \right|^2 + \left| q_{-1} \right|^2
    &\Leftarrow& \mbox{half-integer}\ J,
    \\*[1mm]
    \left| q_0 \right|^2.
    &\Leftarrow& \mathrm{integer}\ J.
  \end{array} \right.
  \label{fffft+3}
  \ee
\end{itemize}
The quartic part of the SP thus is
\be
V_4 =
\frac{\lambda_1}{2}\, F_1^2
+ \frac{\lambda_2}{2}\, F_2^2
+ \lambda_3 F_1 F_2
+ \lambda_4 F_4
+ \sum_{i=5}^{t+3} \lambda_i F_i.
\label{mnb}
\ee

A term with the invariant
\be
F_3 \equiv \sum_{I = - 2 J}^{2 J} \left| c_I \right|^2
\ee
has not been included in $V_4$
because $F_3$ linearly depends on the other invariants.
Indeed,
\be
F_3 + \sum_{i=5}^{t+3} F_i = F_2^2.
\ee

\subsection{BFB conditions}

Let us consider again $\left( \chi \otimes \tilde \chi \right)_\mathbf{3}$
given in Eq.~\eqref{uopc}.
The $SU(2)$-invariant quantity
\be
\left| \left( \chi \otimes \tilde \chi \right)_\mathbf{3} \right|^2
\equiv \left( \sum_{I=-J}^J I \left| \chi_I \right|^2 \right)^2
+ \left| \sum_{I=1-J}^J \chi_I^\ast \chi_{I-1} \sqrt{J^2 - I^2 + J + I} \right|^2
\ee
is four-linear in the $\chi_I$
and therefore it must be linearly dependent on $F_2^2$ and the $F_i$.
Indeed,
one finds that
\be
\label{summ_kappFi}
\left| \left( \chi \otimes \tilde \chi \right)_\mathbf{3} \right|^2
= J^2 F_2^2 - \sum_{i=5}^{t+3} \kappa_i F_i,
\ee
where the numbers $\kappa_i$ are given by
\be
\label{ki}
\kappa_i = \left( i - 4 \right) \left( 4 J + 9 - 2 i \right).
\ee
Notice that all the $\kappa_i$ are \emph{positive}.
We have explicitly checked,
up to $J = 10$,
that Eqs.~\eqref{summ_kappFi} and~\eqref{ki} are correct.

From Eq.~\eqref{summ_kappFi},
\be
\sum_{i=5}^{t+3} \kappa_i F_i - J^2 F_2^2 =
- \left( \sum_{I = - J}^J I \left| \chi_I \right|^2 \right)^2
- \left| \frac{z}{a b} \right|^2,
\ee
where $z$ has been defined in Eq.~\eqref{z};
hence,
from the definition of $F_4$ in Eq.~\eqref{ffff4},
\be
\sum_{i=5}^{t+3} \kappa_i F_i - J^2 F_2^2 =
- \left( \frac{2 F_4 - z - z^\ast}{\left| a \right|^2 - \left| b \right|^2}
\right)^2
- \left| \frac{z}{a b} \right|^2.
\ee
Therefore,
\ba
\sum_{i=5}^{t+3} \kappa_i F_i - J^2 F_2^2 + \frac{4 F_4^2}{F_1^2}
&=& \frac{1}{4 \left| a b \right|^2 \left( \left| a \right|^4
  - \left| b \right|^4 \right)^2} \left\{
\vphantom{\left[ 8 \left| a b \right|^2 F_4
  - \left( \left| a \right|^2 + \left| b \right|^2 \right)^2
  \left( z + z^\ast \right) \right]^2}
\left( \left| a \right|^4 - \left| b \right|^4 \right)^2
\left( z - z^\ast \right)^2
\right. \no & & \left.
- \left[ 8 \left| a b \right|^2 F_4
  - \left( \left| a \right|^2 + \left| b \right|^2 \right)^2
  \left( z + z^\ast \right) \right]^2
\right\}.
\ea
Thus,
\be
\sum_{i=5}^{t+3} \kappa_i F_i - J^2 F_2^2 + \frac{4 F_4^2}{F_1^2} \le 0.
\label{prp}
\ee

We now define the dimensionless quantities~\cite{fonseca}\footnote{The
Klein--Gordon fields $a$,
$b$,
and $\chi_I$ have mass dimension,
hence $\left[ F_1 \right] = \left[ F_2 \right] = M^2$
and $\left[ F_4 \right] = \left[ F_i \right] = M^4$,
for $i = 5, \ldots, t+3$.}
\bs
\ba
r &\equiv& \frac{F_1}{F_2}, \\
\gamma_i &\equiv& \frac{F_i}{F_2^2} \quad (i = 5, \ldots, t+3), \\
\delta &\equiv& \frac{2 F_4}{J F_1 F_2}.
\ea
\label{r_delt_gamm}
\es
We then have,
from Eq.~\eqref{mnb},
\bs
\label{vvvv4}
\ba
\frac{V_4}{F_2^2} &=&
\frac{\lambda_1}{2}\, r^2
+ \frac{\lambda_2}{2}
+ \lambda_3 r
+ \frac{J}{2}\, \lambda_4 r \delta
+ \sum_{i=5}^{t + 3} \lambda_i \gamma_i
\\ &=& \frac{1}{2}
\left( \begin{array}{cc} r, & 1 \end{array} \right)
\left( \begin{array}{cc} \lambda_1 &
  \lambda_3 + \displaystyle{\frac{J}{2}\, \lambda_4 \delta}
  \\*[3mm]
  \lambda_3 + \displaystyle{\frac{J}{2}\, \lambda_4 \delta} &
  \displaystyle{\lambda_2 + 2 \sum_{i=5}^{t+3} \lambda_i \gamma_i}
\end{array} \right)
\left( \begin{array}{c} r \\ 1 \end{array} \right).
\label{u48843}
\ea
\es
It follows from the definitions of $F_2$ and $F_1$
in Eqs.~\eqref{ffff2} and~\eqref{ffff1},
respectively,
that $r \ge 0$.
Therefore,
the conditions for $V_4 \left/ F_2^2 \right.$ in Eq.~\eqref{u48843}
to be non-negative are~\cite{kannike}
\bs
\label{gen_cop}
\ba
\lambda_1 &\ge& 0, \label{condi1} \\
\lambda_2 + 2\, \sum_{i=5}^{t+3} \lambda_i \gamma_i &\ge& 0,
\label{gen_cop1} \\
\lambda_3 + \frac{J}{2}\, \lambda_4 \delta &\ge& - \sqrt{ \lambda_1 \left(
  \lambda_2 +  2\, \sum_{i=5}^{t+3} \lambda_i \gamma_i \right)}.
\label{gen_cop2}
\ea
\es
The conditions~\eqref{gen_cop1} and~\eqref{gen_cop2}
must hold for all possible values of $\delta$ and of the $\gamma_i$.
It follows from the definitions of the $F_i$,
\textit{cf.}\ Eqs.~\eqref{ffff5} and~\eqref{ffff6},
that the $\gamma_i \ge 0$.
From Eq.~\eqref{prp},
\be
\sum_{i=5}^{t+3} \kappa_i \gamma_i - J^2
\left( 1 - \delta^2 \right)
\le 0.
\ee
Thus,
since all the $\kappa_i$ and the $\gamma_i$ are positive,
\be
\label{prp2}
0 \le \sum_{i=5}^{t+3} \kappa_i \gamma_i \le
J^2 \left( 1 - \delta^2 \right),
\ee
and therefore $-1 \le \delta \le +1$.\footnote{We shall implicitly
assume that the conditions $\left| \delta \right| \le 1$
and~\eqref{prp2} completely determine the parameter space,
\textit{i.e.}\ that no further conditions restrict the parameters
$\delta$ and $\gamma_i$.
Equivalently,
we assume that,
for any parameters $F_2 \ge 0$,
$r \ge 0$,
$\delta \in \left[ -1, +1 \right]$,
and $\gamma_i$ obeying condition~\eqref{prp2},
it is possible to find fields $a$,
$b$,
and $\chi_I$ satisfying Eqs.~\eqref{ffff2},
\eqref{ffff1},
\eqref{ffff4}--\eqref{z},
\eqref{ffff5}--\eqref{fffft+3},
and~\eqref{r_delt_gamm}.
We thank Renato Fonseca for calling our attention
to this implicit assumption of our work.}
It is advantageous to define
\be
x = \frac{1 + \delta}{2} \in \left[ 0, 1 \right].
\ee
Then
Eq.~\eqref{prp2} reads
\be
\label{prp3}
0 \le \sum_{i=5}^{t+3} \kappa_i \gamma_i \le
4 J^2 x \left( 1 - x \right).
\ee
This condition determines the domain of the $\gamma_i$,
which has corners at the points~\cite{fonseca}
\bs
\label{maxpoint}
\ba
\gamma_5 = \cdots = \gamma_{t+3} = 0, & &
\label{jvififo} \\
\gamma_5 = \cdots = \gamma_{t+2} = 0,
& & \gamma_{t+3} = \frac{J^2}{\kappa_{t+3}},
\\
\gamma_5 = \cdots = \gamma_{t+1}= \gamma_{t+3} = 0,
& & \gamma_{t+2} = \frac{J^2}{\kappa_{t+2}},
\\
\vdots && \vdots
\no
\gamma_6 = \cdots = \gamma_{t+3} = 0,
& & \gamma_{5} = \frac{J^2}{\kappa_{5}}.
\ea
\es
Since $\lambda_2 +  2 \sum_{i=5}^{t+3} \lambda_i \gamma_i$
is a linear function of the $\gamma_i$,
condition~\eqref{gen_cop1} just has to hold
at the corners of the domain of the $\gamma_i$
in order to hold in the whole domain.
We thus obtain \emph{necessary} BFB conditions:
\bs
\label{BFBnec}
\ba
\lambda_2 &\ge& 0, \label{condlam2} \\
\widehat{\lambda}_i &\ge& 0, \label{uwe}
\ea
\es
where
\bs
\ba
\widehat{\lambda}_i &\equiv& \lambda_2 + q_i,
\\
q_i &\equiv& \frac{2 J^2}{\kappa_i}\, \lambda_i.
\ea
\es
Furthermore,
condition~\eqref{gen_cop2} must certainly hold at the point~\eqref{jvififo}
and for both $\delta = 0$ and $\delta = \pm 1$.
Therefore,
\bs
\label{condlam}
\ba
\lambda_3 &\ge& - \sqrt{ \lambda_1 \lambda_2},
\label{condlam3} \\
\left| \lambda_4 \right| &\le&
\frac{2}{J} \left( \lambda_3 + \sqrt{\lambda_1 \lambda_2} \right).
\label{condlam4}
\ea
\es
The \emph{necessary} BFB conditions~\eqref{condlam3} and~\eqref{condlam4}
generalize conditions~\eqref{www3} and~\eqref{www4},
respectively,
when $\lambda_2$ is nonzero.

Condition~\eqref{gen_cop2}
must also hold at all the other corners~\eqref{maxpoint}
of the $\gamma_i$ domain.
Therefore,
\be
\lambda_3 - \frac{J}{2}\, \lambda_4 + J \lambda_4 x
\ge - \sqrt{ \lambda_1
  \left[ \lambda_2 + 4 q_i x \left( 1 - x \right) \right]}
\label{BFBsuff1}
\ee
must hold for all $i = 5, \ldots, t+3$
and for all $x \equiv \left[ 0, 1 \right]$.
Thus,
the functions
\be
f_i \left( x \right) \equiv
\lambda_3 - \frac{J}{2}\, \lambda_4 + J \lambda_4 x + \sqrt{ \lambda_1 \left[
    \lambda_2 + 4 q_i \left( x - x^2 \right) \right]}
\quad
(i = 5, \ldots, t+3)
\ee
must be non-negative $\forall x \in \left[ 0, 1 \right]$.
Clearly~\cite{fonseca},
\bs
\ba
\frac{\mathrm{d} f_i}{\mathrm{d} x} &=&
J \lambda_4 + \frac{2 \sqrt{\lambda_1}\,
  q_i \left( 1 - 2 x \right)}{\sqrt{\lambda_2 + 4 q_i \left( x - x^2 \right)}},
\\
\frac{\mathrm{d}^2 f_i}{\mathrm{d} x^2} &=&
\frac{- 4 q_i \widehat \lambda_i \sqrt{\lambda_1}}{\left[ \sqrt{\lambda_2
      + 4 q_i \left( x - x^2 \right)} \right]^3}.
\ea
\es
Because of Eq.~\eqref{uwe},
the second derivative of $f_i$ has sign
opposite to the one of $q_i$,
\textit{i.e.}\ opposite to the one of $\lambda_i$.
Since we have already ascertained---through condition~\eqref{condlam4}---that
both $f_i \left( 0 \right) \ge 0$ and $f_i \left( 1 \right) \ge 0$,
the condition $f_i \left( x \right) \ge 0,\ \forall x \in \left[0, 1 \right]$
is equivalent to the following:
\begin{itemize}
\item either $\mathrm{d}^2 f_i \left/ \mathrm{d} x^2 \right. < 0$,
\item or there is no real number $x_0$
  such that $f^\prime \left( x_0 \right) = 0$,
\item or such a $x_0$ exists,
  but it is outside the interval $\left[ 0, 1 \right]$,
\item or $f \left( x_0 \right) \ge 0$.
\end{itemize}
This is equivalent to
\bs
\label{BFBsuff2}
\ba
& & \mbox{either} \quad \lambda_i > 0,
\\
& & \mbox{or} \quad \lambda_i \Lambda_i < 0,
\\
& & \mbox{or} \quad
\sqrt{\frac{\widehat{\lambda}_i}{q_i\, \Lambda_i}}
> \frac{2}{J \left| \lambda_4 \right|},
\\
& & \mbox{or} \quad
\lambda_3 \ge - \sqrt{\frac{\widehat{\lambda}_i\, \Lambda_i}{q_i}},
\ea
\es
respectively,
where
\be
\Lambda_i \equiv \frac{J^2}{4}\, \lambda_4^2 + q_i \lambda_1.
\ee
Conditions~\eqref{condi1},
\eqref{BFBnec},
\eqref{condlam},
and~\eqref{BFBsuff2}
($i = 5, \ldots, t+3$)
are \emph{necessary and sufficient} for
the boundedness-from-below of $V_4$~\cite{fonseca}.

\subsection{UNI conditions}

The condition~\eqref{uniarb2}
stays unchanged when there are in $V_4$ terms four-linear in the $\chi_I$.

The eigenvalues of the scattering matrix of the two-field states
with null $T_3$ and hypercharge $2Y$,
\textit{viz.}\ the states $\chi_J \chi_{-J},\ \chi_{J-1} \chi_{1 - J},\
\chi_{J-2} \chi_{2 - J},\ \ldots$ produce the conditions
\bs
\label{uniarb33}
\ba
\left| \lambda_2 \right| &<& M,
\\
\left| \lambda_2 + 2 \lambda_i \right| &<& M \quad (i = 5, \ldots, t+3).
\ea
\es

The scattering matrix for the two-field states
with null hypercharge and null third component of isospin---\textit{i.e.},
for the states $\left| a \right|^2$,
$\left| b \right|^2$,
and the $n$ states $\left| \chi_I \right|^2$---generalizes
the matrix of Eq.~\eqref{jbogfof}:
\be
\mathcal{S} = \left( \begin{array}{ccc}
  2 \lambda_1 & \lambda_1 & \Sigma_1 \\
  \lambda_1 & 2 \lambda_1 & \Sigma_2 \\
  \Sigma_1^T & \Sigma_2^T & \Lambda
  \end{array} \right)
\label{matrix00}
\ee
where the $1 \times n$ submatrices $\Sigma_1$ and $\Sigma_2$ are given by
\bs
\ba
\left( \Sigma_1 \right)_{1k}
&=& \lambda_3 + \frac{\lambda_4}{2} \left( J + 1 - k \right),
\\
\left( \Sigma_2 \right)_{1k}
&=& \lambda_3 - \frac{\lambda_4}{2}\left( J + 1 - k \right),
\ea
\es
for $k = 1, \ldots, n$.
The $n \times n$ matrix $\Lambda$ is given by
\bs
\ba
\Lambda_{kl} &=& \lambda_2 \left( 1 + \delta_{kl} \right)
\\ & &
+ 4\, \sum_{i=5}^{t+3} \lambda_i\! \sum_{m=1}^{4 J + 17 - 4 i}
\delta_{m,\, k + l + 7 - 2 i}
\no & &
\times \left( \left[ \begin{array}{ccccc}
  J & & J & & 2 J + 8 - 2 i \\ 
  J + 1 - k & & J + 1 - l & & 2 J + 9 - 2 i - m
\end{array} \right] \right)^2,
\ea
\es
for $k, l \in \left[ 1, n \right]$.
For all the values of $J$ that we have investigated
(\textit{i.e.}\ for all integer and half-integer $J$ up to and including 5),
the matrix $\mathcal S$ of Eq.~\eqref{matrix00}
is equivalent to the direct sum of
\begin{itemize}
\item $2 J - 1$ $1 \times 1$ matrices,
  \textit{i.e.}\ numbers that are linear combinations of $\lambda_2$
  and the $\lambda_i$ ($i = 5, \ldots, t+3$),
  the coefficient of $\lambda_2$ in those linear combinations being 1;
\item one $2 \times 2$ symmetric matrix with
  \begin{itemize}
  \item
    $\left( 1, 1 \right)$ matrix element $\lambda_1$,
  \item $\left( 2, 2 \right)$ matrix element which is
    a linear combination of $\lambda_2$ and the $\lambda_i$,
    the coefficient of $\lambda_2$ in that linear combination being 1,
  \item $\left( 1, 2 \right)$ matrix element proportional to $\lambda_4$;
  \end{itemize}
\item another $2 \times 2$ symmetric matrix with
  \begin{itemize}
  \item
    $\left( 1, 1 \right)$ matrix element $3 \lambda_1$,
  \item $\left( 2, 2 \right)$ matrix element which is
    a linear combination of $\lambda_2$ and the $\lambda_i$,
    the coefficient of $\lambda_2$ in that linear combination being $2 J + 2$,
  \item $\left( 1, 2 \right)$ matrix element proportional to $\lambda_3$.
  \end{itemize}
\end{itemize}
The moduli of all the eigenvalues of these $2 J + 1$ matrices
should be smaller than $M$.
Since the matrices are either $1 \times 1$ or $2 \times 2$,
there are simple analytic expressions for their $2 J + 3$ eigenvalues.
In particular,
from the last two matrices mentioned above one obtains the unitarity conditions
\bs
\label{genuni}
\ba
\left| \lambda_1 + A_1 \right| + \sqrt{\left( \lambda_1 - A_1 \right)^2
  + \frac{2}{3}\, J \left( J + 1 \right) \left( 2 J + 1 \right) \lambda_4^2}
&<& 2 M,
\label{genuni2}
\\
\left| 3 \lambda_1 + A_2 \right| + \sqrt{\left( 3 \lambda_1 - A_2 \right)^2
  + 8 \left( 2 J + 1 \right) \lambda_3^2}
&<& 2 M,
\label{genuni1}
\ea
\es
where
\bs
\ba
A_1 &\equiv& \lambda_2 + 4\,
\sum_{i=5}^{t+3}\, 
  \frac{J^2 + 16 J + 36 + i \left( 2 i - 4 J - 17 \right)}{J
  \left( J + 1 \right)}\ \frac{4 J - 4 i + 17}{2 J + 1}\, \lambda_i,
\\
A_2 &\equiv& 2 \left( J + 1 \right) \lambda_2
+ 4\, \sum_{i=5}^{t+3} \frac{4 J - 4 i + 17}{2 J + 1}\, \lambda_i.
\ea
\es
We have explicitly checked that Eqs.~\eqref{genuni} are correct
up to $J = 11/2$.

The unitarity conditions that one
obtains
from the other $2 J$ matrices
are explicitly given in Appendix~\ref{uniconditions} for all $J$ through $7/2$.
We point out that our unitarity conditions for the case $J = 3/2$
do not perfectly coincide with the ones given in Ref.~\cite{kannikerecent}.

So,
the full UNI conditions are\footnote{Condition~\eqref{l1cond}
is unimportant in practice,
because we already know that $\lambda_1 = 0.258$ is quite small.}
\be
\left| \lambda_1 \right| < M,
\label{l1cond}
\ee
condition~\eqref{uniarb2},
conditions~\eqref{uniarb33},
condition~\eqref{genuni},
and the conditions in Appendix~\ref{uniconditions}.

\subsection{Results}
\label{sect:results}

We have generated random sets of values
for all the coefficients of $V_4$ except $\lambda_1$,
\textit{viz.}\ for $\lambda_2$,
$\lambda_3$,
$\lambda_4$,
and the $\lambda_i$
($i = 5, \ldots, t+3$).
The coefficient $\lambda_1$ was kept fixed at the value~\eqref{lambda1}.
We have then imposed on the generated sets both the BFB and the UNI conditions,
thereby discarding most of them.
We have made scatter plots of the sets of values that respected
both the BFB and the UNI conditions.
By carefully scrutinizing those plots,
we have arrived at the maximum and minimum allowed values of $\lambda_3$,
and at the maximum allowed value of $\left| \lambda_4 \right|$,\footnote{The
coefficient $\lambda_4$ can always be zero,
\textit{i.e.}\ the minimum allowed value of $\left| \lambda_4 \right|$ is zero.}
that are displayed in Table~\ref{jcido}.
These values were also checked through a fitting procedure,
by using both the UNI and BFB conditions.
\begin{table}
\centering
\begin{tabular}{||c||c|c|c|c|c|c|c||}
  \hline
  $J$ & $1/2$ & $1$ & $3/2$ & $2$ & $5/2$ & $3$ & $7/2$ \\ \hline
  maximum $\left| \lambda_4 \right|$ & $26.46$ & $17.49$ & $11.96$ &
  $8.10$ & $5.97$ & $4.65$ & $3.76$ \\ \hline
  maximum $\lambda_3$ & $12.37$ & $10.10$ & $8.75$ &
  $7.82$ & $7.14$ & $6.61$ & $6.19$ \\ \hline
  minimum $\lambda_3$ & $-1.46$ & $-1.26$ & $-1.13$ &
  $-1.03$ & $-0.95$ & $-0.89$ & $-0.84$ \\ \hline
\end{tabular} 
\caption{The maximum allowed value of $\left| \lambda_4 \right|$,
  and the maximum and minimum allowed values of $\lambda_3$,
  for various values of $J$.
  \label{jcido}}
\end{table}

It turns out that the maximum value of $\left| \lambda_4 \right|$,
when $\lambda_2$ and the $\lambda_i$ are allowed to be nonzero,
is slightly larger than Eq.~\eqref{IIIlambda4} for $J$ either $1/2$ or $1$,
and slightly larger than Eq.~\eqref{quatro} for all larger values of $J$.
This is illustrated in the left panel of Fig.~\ref{fig:lmd4_solut}.

\begin{figure}[ht]
\begin{center}
\includegraphics[width=0.5\textwidth]{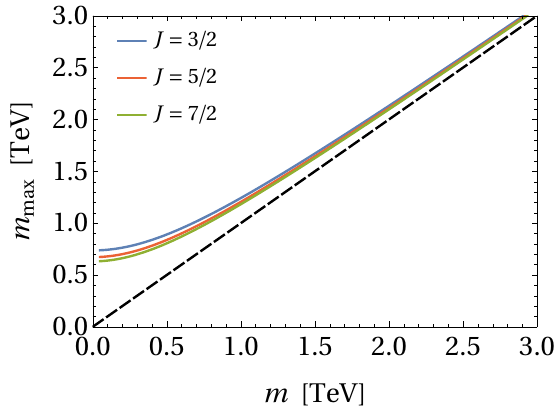}
\end{center}
\caption{The maximal mass $m_\mathrm{max}$ of a scalar multiplet $\chi$
  \textit{versus}\/ its lowest mass $m$,
  for three different values of the isospin $J$ of $\chi$.}
\label{fig:m_max_vs_m}
\end{figure}

In Fig.~\ref{fig:m_max_vs_m} we depict the maximum possible mass
of a multiplet of scalars as a function of its minimum mass $m$.
This is simply given by the expression
\be
m_\mathrm{max} = \sqrt{m^2 + J v^2 \left| \lambda_4 \right|_\mathrm{maximal}},
\ee
where $J$ is the isospin of the multiplet,
\be
v^2 = \frac{1}{2 \sqrt{2}\, G_F} \approx \left( 174\,\mathrm{GeV} \right)^2,
\ee
and $\left| \lambda_4 \right|_\mathrm{maximal}$ is the maximum allowed value
of $\left| \lambda_4 \right|$ for each $J$.
One sees that heavy scalar multiplets tend to be almost degenerate;
for $m \gtrsim 2$~TeV,
$m_\mathrm{max} - m \sim 100$~GeV.
Notice that $m_\mathrm{max} - m$ is maximal for $J = 3/2$,
\textit{i.e.}\/ when $\chi$ is a quadruplet;
if $\chi$ is a larger multiplet,
then it has more components but,
as $\left| \lambda_4 \right|_\mathrm{maximal}$ is smaller,
those components are packed into an ever smaller mass range.

The maximum allowed value of $\lambda_3$
is always attained when $\lambda_2$ and all the $\lambda_i$ vanish,
and exactly coincides with Eq.~\eqref{cio1},
as is illustrated in the right panel
of Fig.~\ref{fig:lmd4_solut}.\footnote{When the coefficient $\lambda_3$
attains its maximum allowed value
displayed in the third row of Table~\ref{jcido},
$\lambda_4$ may have various values,
including zero.}

The minimum allowed value of $\lambda_3$
is always attained when both $\lambda_4$ and all the $\lambda_i$ are zero,
but $\lambda_2$ is nonzero.
Indeed,
the minimum value of $\lambda_3$ is determined by
the BFB condition~\eqref{condlam4}---with
$\lambda_4$ taken to zero---together with the UNI condition
\be
\left| 3 \lambda_1 + 2 \left( 1 + J \right) \lambda_2 \right|
+ \sqrt{\left[ 3 \lambda_1 - 2 \left( 1 + J \right) \lambda_2 \right]^2
  + 8 \left( 1 + 2 J \right) \lambda_3^2} \le 2 M,
\label{kgll}
\ee
which holds when all the $\lambda_i$ are taken to zero.
Thus,
when conditions~\eqref{condlam4} and~\eqref{kgll}
are transformed into equations,
they produce the solution
\bs
\ba
\lambda_2 &=& \frac{1}{2}\,
\frac{M \left( M - 3 \lambda_1 \right)}{\left( 1 + J \right) M
  - \left( 2 + J \right) \lambda_1},
\\
\lambda_3 &=& - \sqrt{\frac{\lambda_1}{2}\
\frac{M \left( M - 3 \lambda_1 \right)}{\left( 1 + J \right) M
  - \left( 2 + J \right) \lambda_1}}. \label{minl3}
\ea
\es
Equation~\eqref{minl3} gives the minimum value of $\lambda_3$
in the fourth row of Table~\ref{jcido}.

\section{Oblique parameters}
\label{ops}

In our NP model it is possible---depending on the values of $J$ and $Y$---that
the new scalars do not couple to the light fermions at all.
If that is so and if,
moreover,
the new scalars are very heavy,
so that they cannot be produced at the LHC---for instance
through the Drell--Yan process---then
they will make themselves felt only indirectly
through their oblique corrections,
\textit{i.e.}~through their contributions
to the self-energies of the gauge bosons.
Following Maksymyk \textit{et al.}~\cite{maksymyk},
we parameterize those corrections through six oblique parameters (OPs) $S$,
$T$,
$U$,
$V$,
$W$,
and $X$.
We use as input of the renormalization process the quantities
$\alpha$ (the fine-structure constant),
$G_F$ (the Fermi coupling constant),
and $m_Z$ (the $Z^0$ boson mass).
Following Ref.~\cite{PDG2022},
we use $\alpha \left( m^2_Z \right) = 1 \! \left/ 127.951 \right.$,
$G_F=1.1663788 \times 10^{-5}$~GeV$^{-2}$,
and $m_Z=91.1876$~GeV.
We then define the weak mixing angle $\theta_W$ through
\be
c_W^2 s_W^2 = \frac{\pi \alpha}{\sqrt{2} G_F m_Z^2},
\ee
where $c_W \equiv \cos{\theta_W}$ and $s_W \equiv \sin{\theta_W}$.
This results in $s_W^2 = 0.23356$.
We then use $S, \ldots, X$ to parameterize,
for each electroweak observable $O$,
the ratio between the prediction of the NP model and the prediction of the SM,
by using expressions of the general form
\be
\frac{O_\mathrm{NP}}{O_\mathrm{SM}} = 1
+ c_S^O S + c_T^O T + c_U^O U + c_V^O V + c_W^O W + c_X^O X,
\label{ratioNP_SM}
\ee
where the coefficients $c_S^O, \ldots, c_X^O$---given for instance
in Ref.~\cite{dudenas}---are known functions of the input quantities.

\subsection{Formulas for the OPs}

According to Ref.~\cite{alberga},
when in the NP Model there is \emph{only one} $SU(2)$ multiplet
of new scalars with weak isospin $J$
and weak hypercharge $Y$,
the parameter $T$ produced by those scalars is given by
\be
T = \frac{G_F}{8 \sqrt{2} \pi^2 \alpha}\
\sum_{I = 1 - J}^J \left( J^2 + J - I^2 + I \right)
\theta_+ \left( m_I^2, m_{I-1}^2 \right),
\ee
where $m_I$ denotes the mass of the scalar with third component of isospin $I$.
The function $\theta_+ \left( x, y \right)$ is defined as
\be
\theta_+ \left( x, y \right) \equiv \left\{ \begin{array}{l}
  \displaystyle{x + y - \frac{2 x y}{x - y}\, \ln{\frac{x}{y}}}
  \Leftarrow x \neq y,
  \\*[2mm]
  0 \Leftarrow x = y.
\end{array} \right.
\ee

The parameters $V$,
$W$,
and $X$ are given by
\bs
\label{uvifp}
\ba
V &=& \frac{G_F m_Z^2}{\sqrt{2} \pi^2 \alpha}\
\sum_{I=-J}^J \left( I c_W^2 - Y s_W^2 \right)^2\,
\rho \left( \frac{m_I^2}{m_Z^2}, \frac{m_I^2}{m_Z^2} \right),
\\
W &=& \frac{1}{4 \pi s_W^2}\, \sum_{I=1-J}^J \left( J^2 + J - I^2 + I \right)
\rho \left( \frac{m_I^2}{m_W^2}, \frac{m_{I-1}^2}{m_W^2} \right),
\\
X &=& - \frac{1}{2 \pi}\, \sum_{I=-J}^J \left( I + Y \right)
\left( I c_W^2 - Y s_W^2 \right)\,
\zeta \left( \frac{m_I^2}{m_Z^2}, \frac{m_I^2}{m_Z^2} \right),
\ea
\es
respectively.
In Eqs.~\eqref{uvifp},
\bs
\label{68}
\ba
\zeta \left( x, y \right) &=&
\frac{11}{36} - \frac{5 \left( x + y \right)}{12}
+ \frac{x y}{3 \left( x - y \right)^2}
+ \frac{\left( x - y \right)^2}{6}
\no & &
+ \left[
  \frac{x^2 - y^2}{4}
  + \frac{\left( y - x \right)^3}{12}
  + \frac{x^2 + y^2}{4 \left( y - x \right)}
  + \frac{x + y}{12 \left( x - y \right)}
  + \frac{x y \left( x + y \right)}{6 \left( y - x \right)^3} 
  \right] \ln{\frac{x}{y}}
\no & &
+ \frac{\Delta \left( x, y \right)}{12}\ f \left( x, y \right),
\\
\rho \left( x, y \right) &=&
\frac{1}{6} - \frac{3 \left( x + y \right)}{4}
+ \frac{\left( x - y \right)^2}{2}
+ \left[
  \frac{\left( y - x \right)^3}{4}
  + \frac{x^2 + y^2}{4 \left( y - x \right)}
  + \frac{x^2 - y^2}{2} \right] \ln{\frac{x}{y}}
\no & &
+ \frac{\left( x - y \right)^2 - x - y}{4}\ f \left( x, y \right),
\ea
\es
for $x \neq y$,
while
\bs
\label{69}
\ba
\zeta \left( x, x \right) &=& \frac{4}{9} - \frac{4 x}{3}
+ \frac{\Delta \left( x, x \right)}{12}\ f \left( x, x \right),
\\
\rho \left( x, x \right) &=& \frac{1}{6} - 2 x
- \frac{x}{2}\ f \left( x, x \right).
\ea
\es
In Eqs.~\eqref{68} and~\eqref{69},
\be
f \left( x, y \right) = \left\{
\begin{array}{l}
  \displaystyle{\sqrt{\Delta \left( x, y \right)}\
    \ln{\frac{x + y - 1 + \sqrt{\Delta \left( x, y \right)}}{x + y
        - 1 - \sqrt{\Delta \left( x, y \right)}}}\,
    \Leftarrow \Delta \left( x, y \right) \ge 0,}
  \\*[4mm]
  \displaystyle{- 2 \sqrt{- \Delta \left( x, y \right)} \left[
    \arctan{\frac{x - y + 1}{\sqrt{- \Delta \left( x, y \right)}}}
    + \left( x \leftrightarrow y \right) \right]
    \Leftarrow \Delta \left( x, y \right) < 0,}
\end{array}
\right.
\ee
where
\be
\Delta \left( x, y \right)
= 1 - 2 \left( x + y \right) + \left( x - y \right)^2.
\ee

For $S$ and $U$ one has
\be
S = S^\prime + S^{\prime \prime}, \quad U = U^\prime + U^{\prime \prime}.
\ee
where
\bs
\ba
S^{\prime \prime} &=& - \frac{2}{\pi}\, \sum_{I=-J}^J
\left( I c_W^2 - Y s_W^2 \right)^2\,
\zeta \left( \frac{m_I^2}{m_Z^2}, \frac{m_I^2}{m_Z^2} \right),
\\
U^{\prime \prime} &=& - S^{\prime \prime} - \frac{1}{\pi}\, \sum_{I=1-J}^J
\left( J^2 + J - I^2 + I \right)
\zeta \left( \frac{m_I^2}{m_W^2}, \frac{m_{I-1}^2}{m_W^2} \right),
\ea
\es
and
\bs
\ba
S^\prime &=& - \frac{Y}{3 \pi}\,
\sum_{I=-J}^J I\, \ln{\frac{m_I^2}{\mu^2}},
\\
U^\prime &=& \frac{1}{12 \pi}\, \sum_{I=1-J}^J
\left( J^2 + J - I^2 + I \right)
g \left( \frac{m_I^2}{m_{I-1}^2} \right)
+ \frac{1}{6 \pi}\, \sum_{I=-J}^J
\left( J^2 + J - 3 I^2 \right) \ln{\frac{m_I^2}{\mu^2}}. \hspace*{10mm}
\label{kfpfp}
\ea
\es
In Eq.~\eqref{kfpfp},
\be
g \left( x \right) = \left\{ \begin{array}{l}
  {\displaystyle \frac{x^3 - 3 x^2 - 3 x + 1}{\left( x - 1 \right)^3}\, \ln{x}
    - \frac{5 x^2 - 22 x + 5}{3 \left( x - 1 \right)^2}}\
  \Leftarrow\ x \neq 1,
  \\*[2mm]
  0\ \Leftarrow\ x = 1
\end{array} \right.
\ee
is a function that obeys $g \left( x \right) = g \left( 1/x \right)$.

We note that the expresions for the OPs are invariant under
the transformation $I \to -I,\ Y \to -Y$.
This allows one to choose the scalar with $I=-J$ to be the lightest one,
provided one keeps $Y$ free,
\textit{i.e.}~provided one considers both negative and positive values of $Y$;
that is the procedure that we adopt.

\subsection{Numerical results}
\label{sect:NumRes}

In our numerical work we utilize the set of electroweak observables
given in Table~\ref{obs_table}.
\begin{table}[]
\centering
\begin{tabular}{l |+|+|+}
  Observable & \multicolumn{1}{c|}{Measurement ($O_\mathrm{meas}$)} &
  \multicolumn{1}{c|}{SM prediction ($O_\mathrm{SM}$)} &
  \multicolumn{1}{c}{$O_\mathrm{meas}/O_\mathrm{SM}$}
\\ \hline
$\sigma^0_{\textrm{had}}$~[nb] 	&
41.481 + 0.033  & 41.482 + 0.008 & 0.999976 + 0.0008186 \\
$R_\ell$					&
20.767 + 0.025 	& 20.736 + 0.010 	& 1.00149 + 0.001299 \\
$R_b$					&
0.21629 + 0.00066 	& 0.21582 + 0.00002 	& 1.00218 + 0.003060 \\
$R_c$ 					&
0.1721 + 0.0030 	& 0.17221 + 0.00003 	& 0.999361 + 0.01742 \\
$A_{FB}^{(0, \ell)}$ 	&
0.0171 + 0.001		& 0.01617 + 0.00007 	& 1.05751 + 0.06201 \\
$A_{FB}^{(0, b)}$ 		&
0.0996 + 0.0016 	& 0.1029 + 0.0002 	& 0.967930 + 0.01566 \\
$A_{FB}^{(0, c)}$ 		&
0.0707 + 0.0035 	& 0.0735 + 0.0002 	& 0.961905 + 0.04769 \\
$A_{\ell}$ 				&
0.1513 + 0.0021 	& 0.1468 + 0.0003 	& 1.03065 + 0.01446 \\
$A_b$ 					&
0.923 + 0.020 		& 0.9347 	& 0.987483 + 0.02140 \\
$A_c$ 					&
0.670 + 0.027 		& 0.6677 + 0.0001 	& 1.00344 + 0.04044 \\
$\bar{s}_\ell^2~(\textrm{LEP-1})$  &
0.2324 + 0.0012 	& 0.23155 + 0.00004 	& 1.00367 + 0.005185 \\
$\bar{s}_\ell^2~(\textrm{Tevt.})$	 &
0.23148 + 0.00033 & 	0.23155 + 0.00004	& 0.999698 + 0.001436 \\
$\bar{s}_\ell^2~(\textrm{LHC})$  	 &
0.23129 + 0.00033 & 	0.23155 + 0.00004	& 0.998877 + 0.001436 \\
$m_W$~[GeV] 				&
80.377 + 0.012 	& 80.360 + 0.006 	& 1.00021 + 0.0001670 \\
$\Gamma_W$~[GeV] 		&
2.046 + 0.049 		& 2.089 + 0.001 	& 0.979416 + 0.02346 \\
$\Gamma_Z$~[GeV] 		&
2.4955 + 0.0023 	& 2.4941 + 0.0009 	& 1.00056 + 0.0009903 \\
$g_V^{\nu e}$			&
-0.040 + 0.015 	& -0.0397 + 0.0001 	& 1.00756 + 0.37784 \\
$g_A^{\nu e}$			&
-0.507 + 0.014 	& -0.5064 			& 1.00118 + 0.02765 \\
$Q_W (\mathrm{Cs})$		&
-72.82 + 0.42 	& -73.24 + 0.01 	& 0.994265 + 0.005736 \\
$Q_W (\mathrm{Tl})$		&
-116.4 + 3.6 		& -116.90 + 0.02 	& 0.995723 + 0.03080 \\
\hline
\end{tabular}
\caption{First column: the electroweak observables used in our work.
  Second column: their experimental values, taken from Ref.~\cite{PDG2022}.
  Third column: the SM predictions for them.
  Fourth colum: the ratio between the experimental value and the SM prediction.}
\label{obs_table}
\end{table}
For each set of OPs and for each observable $O$,
we have computed $\left. O_\mathrm{NP} \right/ O_\mathrm{SM}$
by using Eq.~\eqref{ratioNP_SM}.
We have then computed the residuals,
defined as $\left. O_\mathrm{NP} \right/ O_\mathrm{SM}$
minus the values in the last column of Table~\ref{obs_table}.
The $\chi^2$ function for each set of OPs was defined as $\chi^2 = R C^{-1} R^T$,
where $R$ is the row-vector of the residuals and $C$ is the covariance matrix;
the latter is evaluated according to the correlations
among the observables~\cite{PDG2022,ALEPH,Tenchini}.

For each set of OPs,
the pull is evaluated as $\left. r \right/ \delta^\pm$,
where $r$ is the residual defined above and $\delta^\pm$ 
is the error given in the fourth column of Table~\ref{obs_table}.

Firstly,
setting $V = W = X = 0$ and freely adjusting $S$,
$T$,
and $U$ we have accomplished our best fit
of the electroweak observables in Table~\ref{obs_table}.
We have obtained $\chi^2 = 14.201$ for $S = -1.2 \times 10^{-2}$,
$T = 2.8 \times 10^{-2}$,
and $U = 2.0 \times 10^{-3}$.

In our NP model,
for each value of the isospin $J$ of the multiplet,
there are just three free parameters:
\begin{itemize}
\item $\left| \lambda_4 \right|$,
  which determines the mass-squared difference $\Delta m^2
  = \left. \left| \lambda_4 \right| v^2 \right/ 2$
  between any two successive components of the multiplet.
\item The mass $m$ of the lightest component of the multiplet;
  without loss of generality we take that component to be the one
  with the smallest third projection of isospin.
  Thus,
  $m_I^2 = m^2 + \left( I + J \right) \Delta m^2$
  for $I = -J, \ldots, $.
\item The hypercharge $Y$ of the multiplet.
\end{itemize}
For instance,
by choosing $J = 2$,
$m = 3$~TeV,
$\lambda_4 = 3.65$,
and $Y = 1.65$ we have obtained $\chi^2 = 14.2015$, 
which is not very far from our best fit.
We thus see that our model is able to fit the electroweak observables
just as perfectly as a free fit.

For each value of $J$ up to $7/2$---the upper bound on $J$
found in Ref.~\cite{logan}---we let $Y$ vary
from $-Y_\mathrm{max}$ to $Y_\mathrm{max}$,
where $Y_\mathrm{max}$ is the $J$-dependent
upper bound on $\left| Y \right|$ determined in Refs.~\cite{logan,milagre}.
We let $m$ vary from 50~GeV to 3~TeV,
and we let $\left| \lambda_4 \right|$ vary from zero
to its maximum allowed value given in Table~\ref{jcido}.
We keep only the points that either
\begin{enumerate}
\item have $\chi^2$ smaller than 30 \emph{and} all the pulls
  smaller (in modulus) than three,
\item or have $\chi^2$ smaller than 17 and \emph{all}
  the pulls smaller than one,
  \emph{except, possibly},
  the pulls of $A_{FB}^{\left( 0, b \right)}$,
  $A_\ell$,
  $R_\ell$,
  and $Q_W\!\left( \mathrm{Cs} \right)$.
\end{enumerate}
In this way we obtain two sets of points,
that we use to construct Figures~\ref{fig:lmd4_vs_m} and~\ref{fig:lmd4_vs_Y}
below.
Most pulls of the observables are always very small;
only a few observables have large pulls.
As a consequence,
in practice,
points with $\chi^2 < 30$ mostly have all the pulls between $-3$ and $+3$,
and points with $\chi^2 < 17$ almost always
have all the pulls between $-1$ and $+1$,
except for the observables $A_{FB}^{\left( 0, b \right)}$
$A_\ell$,
$R_\ell$,
and $Q_W\!\left( \mathrm{Cs} \right)$.\footnote{We make the exception
of $Q_W\!\left(\mathrm{Cs}\right)$ because,
if one forces its pulls to be smaller than one,
that noticeably restricts the parameter space,
by practically eliminating all the negative values of $Y$.}

One sees in Fig.~\ref{fig:lmd4_vs_m} that,
unless $m$ is very large and,
therefore,
the OPs are very small,
the restrictions on $\left| \lambda_4 \right|$ from the OPs
are usually stronger that the UNI and BFB conditions
that we have derived in this paper.
Indeed,
for $\chi^2 \le 30$ and $m \lesssim 2$~TeV
the restrictions from the OPs are stronger,
and the same happens for $\chi^2 \le 17$ and $m \lesssim 3$~TeV.
\begin{figure}[ht]
\begin{center}
\includegraphics[width=1.0\textwidth]{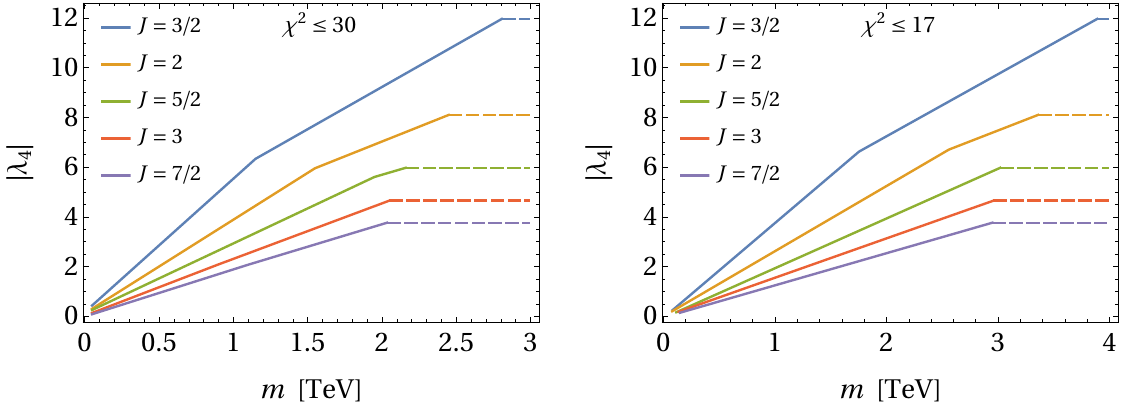}
\end{center}
\caption{The maximum allowed value of $\left| \lambda_4 \right|$
  \textit{versus} the lightest mass $m$,
  for various values of $J$ and for fits with $\chi^2 \le 30$ (left panel)
  or $\chi^2 \le 17$ (right panel).
  The hypercharge $Y$ was left free.
  The horizontal dashed lines correspond to the bounds
  on $\left| \lambda_4 \right|$ from the UNI and BFB conditions.}
\label{fig:lmd4_vs_m}
\end{figure} 

The relation between the upper bound on $\left| \lambda_4 \right|$
and the hypercharge $Y$ is quite complex and very much depends on $m$
(because,
if $m$ gets larger,
then the OPs get smaller and therefore the OPs do not constrain
$\left| \lambda_4 \right|$).
In Fig.~\ref{fig:lmd4_vs_Y},
which was made for $m = 3$~TeV,
one observes that,
as $Y$ increases,
the upper bound on $\left| \lambda_4 \right|$ slightly decreases.
\begin{figure}[ht]
\begin{center}
\includegraphics[width=1.0\textwidth]{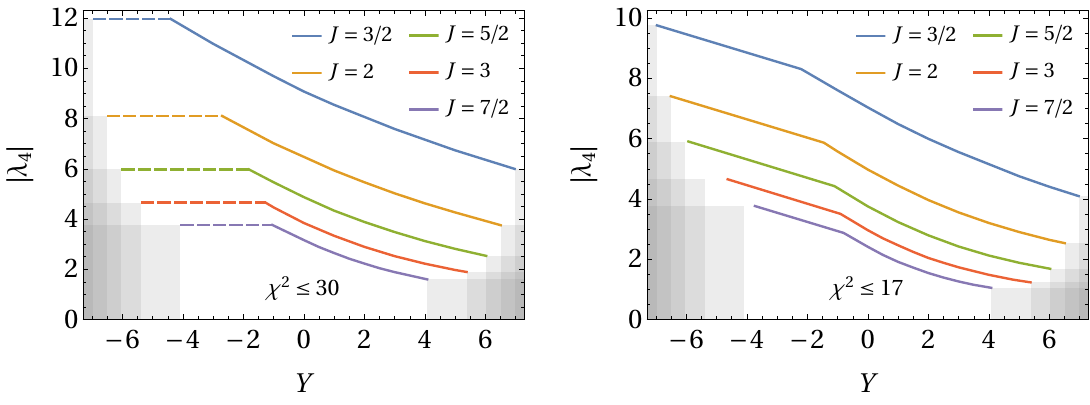}
\end{center}
\caption{The upper bound on $\left| \lambda_4 \right|$
  \textit{versus} the hypercharge $Y$,
  for various values of $J$,
  for $m = 3$~TeV,
  and for fits with $\chi^2 \le 30$ (left panel)
  or $\chi^2 \le 17$ (right panel).
  The horizontal dashed lines indicate the upper bounds
  from the UNI and BFB conditions,
  and the curved lines indicate the upper bounds from the OPs.
  The gray bands indicate the $J$-dependent restrictions on $Y$
  derived in Refs.~\cite{logan,milagre}.}
\label{fig:lmd4_vs_Y}
\end{figure} 
If one requires a smaller $\chi^2$ in the fit of the oblique parameters,
then the constraint on $\left| \lambda_4 \right|$ derived therefrom
becomes stronger and eventually,
as one sees in the right panel of Fig.~\ref{fig:lmd4_vs_Y},
the UNI+BFB bound becomes completely ineffective.

\section{Conclusions and outlook}
\label{conclusions}

In this paper we have studied the extension of the SM
through a scalar multiplet $\chi$ with arbitrary isospin $J$
and hypercharge $Y$.
For every value of $J$,
we have included in the scalar potential (SP)
just those terms that are present there for any value of $Y$.
We have especially concentrated on the term~\eqref{coup},
which fixes the squared-mass difference $\Delta m^2$
between the successive components of $\chi$,
\textit{cf.}~Eq.~\eqref{deltam}.
We have derived an upper bound on $\left| \lambda_4 \right|$,
hence on $\Delta m^2$,
from both the unitarity (UNI) and bounded-from-below (BFB) conditions
on the SP.
We have found that,
remarkably,
that upper bound depends crucially,
not just on the UNI conditions,
but also on the BFB ones.
For instance,
the upper bound that we have found is quite more stringent
than the one utilized in the recent Ref.~\cite{chinese2},
which used only UNI conditions.

Remarkably,
we have been able to derive \emph{necessary and sufficient} BFB conditions
on this model,
even when we allowed the presence in the SP
of the most general terms four-linear in the components of $\chi$.
It so happens that those terms,
even if they are quite complicated to account for,
end up relaxing only a little bit the upper bound on $\left| \lambda_4 \right|$,
\textit{cf.}~Fig.~\ref{fig:lmd4_solut}.

Phenomenologically,
the model that we have studied is,
by itself alone,
of little value,
because,
since we have left $Y$ arbitrary,
the multiplet $\chi$ does not have Yukawa couplings to any fermions.
Moreover,
its lightest component is,
for arbitrary $Y$,
electrically charged and,
moreover,
absolutely stable,
which is of course incompatible with observation.
Therefore,
our study can only be understood as a step towards the understanding
of more specific models,
that will have precise values of $J$ and $Y$,
and probably also extra terms in the SP,
\textit{viz.}~higher-dimensional terms.

\vspace*{5mm}

\paragraph{Acknowledgements:}
L.L.\ thanks Kristjan Kannike for a discussion on Ref.~\cite{kannikerecent}
and for having called his attention to Ref.~\cite{fonseca}.
D.J.\ thanks Anders Eller Thomsen
for a discussion about the package {\tt RGBeta};
he also thanks Simonas Drauk\v{s}as for help with {\tt SARAH}.
The work of D.J.\ was supported by the Lithuanian Particle Physics Consortium.
The work of L.L.\ was supported by the Portuguese
Foundation for Science and Technology through projects UIDB/00777/2020,
UIDP/00777/2020,
CERN/FIS-PAR/0002/2021,
and CERN/FIS-PAR/0019/2021.

\newpage

\begin{appendix}

\section{Explicit UNI conditions}
\label{uniconditions}

\setcounter{equation}{0}
\renewcommand{\theequation}{A\arabic{equation}}

For $J$ through $7/2$,
the unitarity conditions that originate in the matrix $\mathcal{S}$
of Eq.~\eqref{matrix00} are,
besides Eqs.~\eqref{genuni},
the following:
\begin{itemize}
\item For $J = 1$,
  \be
  \left| \lambda_2 + \frac{4}{3}\, \lambda_5 \right| < M.
  \ee
\item For $J = 3/2$,
  \bs
  \label{3/2}
  \ba
  \left| \lambda_2 + \frac{3}{5}\, \lambda_5 \right| &<& M,
  \\
  \left| \lambda_2 + \frac{9}{5}\, \lambda_5 \right| &<& M.
  \ea
  \es
\item For $J = 2$,
  \bs
  \ba
  \left| \lambda_2 + \frac{16}{7}\, \lambda_5 - \frac{4}{5}\, \lambda_6
  \right| &<& M,
  \\
  \left| \lambda_2 + \frac{8}{7}\, \lambda_5 + \frac{4}{5}\, \lambda_6
  \right| &<& M,
  \\
  \left| \lambda_2 - \frac{6}{7}\, \lambda_5 + \frac{4}{5}\, \lambda_6
  \right| &<& M.
  \ea
  \es
\item For $J = 5/2$,
  \bs
  \ba
  \left| \lambda_2 + \frac{79}{45}\, \lambda_5 - \frac{22}{35}\, \lambda_6
  \right| &<& M,
  \\
  \left| \lambda_2 + \frac{19}{9}\, \lambda_5 - \frac{2}{7}\, \lambda_6
  \right| &<& M,
  \\
  \left| \lambda_2 + \frac{5}{9}\, \lambda_5 + \frac{10}{7}\, \lambda_6
  \right| &<& M,
  \\
  \left| \lambda_2 - \frac{29}{15}\, \lambda_5 + \frac{46}{35}\, \lambda_6
  \right| &<& M.
  \ea
  \es
\item For $J = 3$,
  \bs
  \ba
  \left| \lambda_2 + \frac{102}{77}\, \lambda_5 + \frac{25}{21}\, \lambda_6
  - \frac{4}{7}\, \lambda_7
  \right| &<& M,
  \\
  \left| \lambda_2 + \frac{18}{77}\, \lambda_5 + \frac{25}{21}\, \lambda_6
  + \frac{4}{7}\, \lambda_7
  \right| &<& M,
  \\
  \left| \lambda_2 + \frac{6}{7}\, \lambda_5 + \frac{10}{21}\, \lambda_6
  - \frac{4}{7}\, \lambda_7
  \right| &<& M,
  \\
  \left| \lambda_2 - \frac{18}{7}\, \lambda_5 + \frac{19}{21}\, \lambda_6
  + \frac{4}{7}\, \lambda_7
  \right| &<& M,
  \\
  \left| \lambda_2 + \frac{194}{77}\, \lambda_5 - \frac{10}{7}\, \lambda_6
  + \frac{4}{7}\, \lambda_7
  \right| &<& M.
  \ea
  \es
\item For $J = 7/2$,
  \bs
  \ba
  \left| \lambda_2 - \frac{1}{14}\, \lambda_5 + \frac{31}{22}\, \lambda_6
  - \frac{13}{14}\, \lambda_7
  \right| &<& M,
  \\
  \left| \lambda_2 + \frac{53}{78}\, \lambda_5 + \frac{119}{66}\, \lambda_6
  - \frac{1}{2}\, \lambda_7
  \right| &<& M,
  \\
  \left| \lambda_2 + \frac{363}{182}\, \lambda_5 - \frac{1}{22}\, \lambda_6
  - \frac{1}{14}\, \lambda_7
  \right| &<& M,
  \\
  \left| \lambda_2 + \frac{7}{78}\, \lambda_5 + \frac{49}{66}\, \lambda_6
  + \frac{7}{6}\, \lambda_7
  \right| &<& M,
  \\
  \left| \lambda_2 - \frac{121}{42}\, \lambda_5 + \frac{1}{6}\, \lambda_6
  + \frac{17}{14}\, \lambda_7
  \right| &<& M,
  \\
  \left| \lambda_2 + \frac{103}{66}\, \lambda_5 - \frac{101}{66}\, \lambda_6
  + \frac{23}{42}\, \lambda_7
  \right| &<& M.
  \ea
  \es
\end{itemize}

\end{appendix}

\end{document}